\documentclass[fleqn,usenatbib]{mnras}

\usepackage{newtxtext,newtxmath}
\usepackage[T1]{fontenc}
\usepackage{ae,aecompl}
\usepackage{algorithm}
\usepackage[noend]{algpseudocode}
\usepackage{hyperref}
\usepackage{color}
\usepackage{float}
\usepackage{placeins}
\usepackage{graphicx}	
\usepackage{amsmath}	
\usepackage{amssymb}	
\usepackage{caption}

\newcommand{\EE}[1]{\times 10^{#1}}
\newcommand{\LJ}{L1-21J}
\newcommand{\timeanddate}{{\tt timeanddate}}
\newcommand{\vimp}{v_{\rm imp}}
\newcommand{\beq}[1]{\begin{equation}\label{#1}}
\newcommand{\eeq}{\end{equation}}

\newcommand{\checked}[1]{\textcolor{black}{#1}}

\newcommand{\LaLoma}{{\it la Loma}}
\newcommand{\RA}{$\alpha$}
\newcommand{\DEC}{$\delta$}

\newcommand{\hl}[1]{\textcolor{black}{#1}}
\newcommand{\hll}[1]{\textcolor{black}{#1}}

\newcommand{\vimpvalue}{14^{+7}_{-6}} 
\newcommand{\thetamax}{38.2} 
\newcommand{\tflash}{0.3} 
\newcommand{\Eimp}{0.3-1.9} 
\newcommand{\Emed}{0.8} 
\newcommand{\Mrange}{8-99} 
\newcommand{\Drange}{19-44} 
\newcommand{\craterRange}{6-13} 
\newcommand{\Mmed}{27} 
\newcommand{\Dmed}{29} 
\newcommand{\cratermed}{9} 
\newcommand{\flashtime}{04:41:38 UTC}
\newcommand{\etarange}{0.001-0.004} 
\newcommand{\logeta}{-2.6\pm0.1} 
\newcommand{\latvisual}{-68.17} 
\newcommand{\lonvisual}{-29.43} 
\newcommand{\latgeo}{-29.43^{+0.30}_{-0.21}} 
\newcommand{\longeo}{-67.89^{+0.07}_{-0.09}} 
\newcommand{\latpar}{-29.42\pm 0.20} 
\newcommand{\lonpar}{-67.90\pm 0.08} 
\newcommand{\RArad}{3.1} 
\newcommand{\DECrad}{-23.7} 
\newcommand{\Gf}{6.7 \pm 0.3} 
\newcommand{\bG}{2.5\EE{-11}} 
\newcommand{\Go}{0.03} 
\newcommand{\deltaLambda}{420.360} 
\newcommand{\fany}{3} 
\newcommand{\logEr}{6.9\pm 0.4} 
\newcommand{\logK}{9.5\pm 0.4} 
\newcommand{\rhoRegolith}{1600} 
\newcommand{\rhoMin}{1000} 
\newcommand{\rhoMax}{3700} 
\newcommand{\rhoComet}{1000} 
\newcommand{\pComet}{1.0} 
\newcommand{\rhoRocky}{2500} 
\newcommand{\pRocky}{59} 
\newcommand{\rhoMetal}{3700} 
\newcommand{\pMetal}{40} 
\newcommand{\RAgeo}{8.17992} 
\newcommand{\DECgeo}{20.25050} 
\newcommand{\Dimpgeo}{356553} 
\newcommand{\RAcen}{8.16674} 
\newcommand{\DECcen}{20.43615} 
\newcommand{\Dcengeo}{357046} 
\newcommand{\RAcenNASA}{8.16604} 
\newcommand{\DECcenNASA}{20.43654} 
\newcommand{\DcenNASA}{357745} 

\title[Meteoroid impacting the Moon 21/01/2019]{Location, orbit and energy of a meteoroid impacting the Moon during the Lunar Eclipse of January 21, 2019}

\author[Zuluaga et al.]{
J.I. Zuluaga,$^{1,3}$
M. Tangmatitham,$^{2}$
P. Cuartas-Restrepo;$^{1,3}$\thanks{Corresponding author: pablo.cuartas@udea.edu.co}
\newauthor \ J. Ospina,$^{3}$
F. Pichardo,$^{5}$
S.A. L\'opez;$^{3}$
K. Pe\~na,$^{5}$
J.M. Gaviria-Posada$^{4}$
\\ 
$^{1}$Solar, Earth and Planetary Physics - SEAP,  Institute of Physics, University of Antioquia, \\ 
\ Calle 70 No. 52-21, Medell\'in, Colombia\\
$^{2}$Department of Physics, Michigan Technological University, Houghton, MI, USA\\
$^{3}$Sociedad Antioque\~na de Astronom\'ia, CAMO \& Orion groups, Medell\'in, Colombia\\
$^{4}$Observatorio la Loma, V\'ia Concepci\'on-San Vicente Ferrer, Colombia\\
$^{5}$Sociedad Astron\'omica Dominicana, Avenida M\'aximo G\'omez esquina C\'esar Nicol\'as Penson, \\ 
\ Plaza de la Cultura, Santo Domingo, Rep\'ublica Dominicana
}

\date{Accepted XXX. Received YYY; in original form ZZZ}

\pubyear{2019}

\begin{document}
\label{firstpage}
\pagerange{\pageref{firstpage}--\pageref{lastpage}}
\maketitle

\begin{abstract}
During lunar eclipse of January 21, 2019 a meteoroid impacted the Moon producing a visible light flash. The impact was witnessed by casual observers offering an opportunity to study the phenomenon from multiple geographical locations. We use images and videos collected by observers in 7 countries to estimate the location, impact parameters (speed and incoming direction) and energy of the meteoroid. Using parallax, we achieve determining the impact location at lat.$\latgeo$, lon.$\longeo$ and geocentric distance \hl{as} $\Dimpgeo$ km. After devising and applying a photo-metric procedure for measuring flash standard magnitudes in multiple RGB images having different exposure times, we found that the flash, had an average G-magnitude $\langle G\rangle = \Gf$. We use {\em gravitational ray tracing} (GRT) to estimate the orbital properties and likely radiant of the impactor.  We find that the meteoroid impacted the moon with a speed of $\vimpvalue$ km/s (70\% C.L.) and at a shallow angle, $\theta<\thetamax$ degrees. Assuming a normal error for our estimated flash brightness, educated priors for the luminous efficiency and object density, and using the GRT-computed probability distributions of impact speed and incoming directions, we calculate posterior probability distributions for the kinetic energy (median $K_{\rm med}=\Emed$ kton), body mass ($M_{\rm med}=\Mmed$ kg) and diameter ($d_{\rm med}=\Dmed$ cm), and crater size ($D_{\rm med}=\cratermed$ m). If our assumptions are correct, the crater left by the impact could be detectable by prospecting lunar probes. These results arose from a timely collaboration between professional and amateur astronomers which highlight the potential importance of citizen science in astronomy.

\end{abstract}




\section{Introduction}
\label{sec:introduction}

In January 21, 2019 the only total lunar eclipse of 2019 took place. Thousands, if not millions of observers, followed the event in America, north Africa and in most of Europe.  As usual, several amateur and professional observatories around the world streamed the whole eclipse over the internet.

A few minutes after the beginning of the total phase of the eclipse, several sources on the internet claimed the observation of a short light flash on the east side of the eclipsed moon. A few hours after, the flash was fully confirmed by the {\it Moon Impacts Detection and Analysis System, MIDAS} \citep{Madiedo2010} in Spain.  According to MIDAS, one meteoroid (hereafter {\it L1-21J}) impacted the darker side of the eclipsed moon at \checked{\flashtime} \citep{Madiedo2019}. In the days after the eclipse, the Royal Observatory\footnote{\url{https://www.rmg.co.uk}} reported a second flash just two minutes after \LJ\ occurring on the western and much brighter limb of the eclipsed moon. To the date of writing, however, this second flash has not been confirmed by other observers, and therefore, it could also be attributable to other effects, such as instrumental artefacts or cosmic rays (see eg. \citealt{Suggs2011}, \citealt{Suggs2014})

Right after the confirmation by MIDAS of the impact, several observers around the world reported the independent detection of the light flash in their own footage.  Although lunar impacts are relatively common,
the impact of  January 21, 2019 is the first one to be detected simultaneously by thousands of observers during a total lunar eclipse. This offers unique opportunities to study this phenomena from different geographical locations, and using different instruments and independent methods from those used by lunar flash surveys (see Section \ref{sec:impacts}).

Here, we present a scientific analysis of the \LJ\ event using observations gathered, independently, by amateur and professional astronomers, in Colombia, the Dominican Republic, USA, Canary Islands, Cape Verde, Czech Republic, Austria, and Germany (see section \ref{sec:data}).
First, we briefly review what is known about impacts by small meteoroids on the Moon (Section \ref{sec:impacts}). Then, we describe the instruments and data we gather and analyse for this work (Section \ref{sec:impacts}). One of the most interesting characteristic of our approach, is the numerical reconstruction of the meteoroid trajectory, which is required to estimate the speed and the incident angle.  For this purpose we use the novel Gravitational Ray Tracing (GRT) technique (Section \ref{sec:orbit}). \hll{Photometric} analysis of our footage provide us estimations of the total energy involved in the impact (Section \ref{sec:energy}); from there we can estimate the {\it posterior probability distribution} (ppd) of the mass and size of the impactor (Section \ref{sec:sizes}). The precise location of the impact and the crater diameter left by the event are also estimated.

\section{Observation of Moon impacts}
\label{sec:impacts}

Impacts on the Earth-Moon system are relatively common (\citealt{Sigismondi00, Sigismondi00a,Newkum2001,Ivanov2001,Ivanov2006,Gallant2009,Moorhead2017,Drolshagen2017} \hll{and \citealt{Silber2018} and references there in}).
\hl{\cite{Drolshagen2017} for instance (see their Figure 5)} estimate that \hll{$\sim10^4-10^5$ small}, low-mass meteoroids (\hll{$\lesssim0.1$ m diameter, $\lesssim 1$ kg mass}) enter into the Earth's atmosphere per year (\hll{$\sim 1-10$} impact per hour).  The Earth/Moon ratio of meteoroid fluxes is estimated to be 1.38 \citep{Ivanov2006}. Therefore, the rate of impacts on the Moon is on a similar order of magnitude.  However, since our satellite lacks a dense atmosphere, the effects of those impacts on its surface are more dramatic and easier to detect.

With the exception of the event described
in the chronicles of Gervase of Canterbury in 1178 \citep{Hartung93} (whose nature is still debated), the visual observation of impacts on the Moon is not very common. Those impacts could be observed under three favourable conditions: 1) in the days close to the new moon when the dark side is illuminated by the planet shine, 2) far from the dark limb, close to first or last quarter and 3) the most favourable but far less frequent condition, during a total lunar eclipse. In fact, the first impact recorded during a total lunar eclipse was probably the one reported by \cite{Sigismondi00} on January 21, 2000 (exactly one metonic cycle ago).

The probability of observing the impact of small body during a total lunar eclipse, is not negligible.  If we assume that, on average, $>0.4$ light flashes having peak magnitudes $<9$, happens in the moon every hour \citep{Suggs2014}, the probability of observing at least one during the totality ($\sim$ 1 hour) is 33\%; the probability of observing exactly two or more is 6\%, etc.  Naturally, most of those impacts will be very dim and hard to detect with small equipment.

In recent years, improved optical and electronic astronomical equipment and prospecting lunar satellites, have allowed the detection of hundreds of ``fresh'' impacts on the moon using two methods: 1) a local method, involving the  repeated observation of the same portions of the lunar surface at different times, from prospecting satellites; and 2) a remote method, which relies on the observation of the short visible light flashes produced during the impacts.

The NASA Lunar Reconnaissance Orbiter (LRO) has successfully tested the first  method\footnote{\url{http://target.lroc.asu.edu/output/lroc/lroc_page.html}}. During a 6 years mission \citep{Keller2016} the LRO has taken high-resolution images (down to 1 meter per pixel) of $70\%$ of the Moon surface, with almost $3\%$ of the surface observed at least two times. During that time, the spacecraft has detected signatures of hundreds of fresh impacts. The present resolution of LRO allows the detection craters as small as 10 m \citep{Speyerer2016}.  LRO fresh impact signatures have been used for calibrating the Moon cratering flux and to test theoretical estimations of  meteoroid fluxes on the Earth-Moon system \citep{Keller2016}.

Particularly famous are two impacts that were first observed from the Earth and afterwards, their associated crater discovered by LRO. The first one was a extremely bright impact happening on March 17, 2013 \citep{Suggs2014}; the second one happened on September 11, 2013 and it was first identified by the MIDAS system \citealt{Madiedo2014} and then observed by LRO (see below).

In the last two decades, several observing systems were designed and built to monitor the Moon, looking-for flash events \citep{Ortiz2000,Ortiz2002, Ortiz2006, Suggs2014, Ortiz2015, Madiedo2015a, Madiedo2015b, Yanagisawa2002}.
The first lunar-flash monitoring program, MIDAS, was established almost two decades ago in Spain \citep{Madiedo2010}.  During that time, MIDAS has detected a significant number of flashes on the Moon, and the data collected have been used to study the population properties of major meteor showers \citep{Madiedo2014,Ortiz2015, Madiedo2015a, Madiedo2015b}. MIDAS was the first of such monitor system confirming the \LJ\ event. In 2006, the NASA Marshall Space Flight Centre started their own monitoring program\footnote{\url{https://www.nasa.gov/centers/marshall/news/lunar/overview.html}}.  To the date, the NASA's system has independently detected hundreds of events and helped to constrain the rate of meteors falling onto the Moon and, in general, the density and flux of meteoroids around the Earth-Moon system \citep{Suggs2011,Suggs2014}.
More recently, the NEO Lunar Impacts and Optical TrAnsients, NELIOTA saw the first light \citep{Xilouris2018}. To date, and after just a few months of operations, at least 55 flashes have been observed by the NELIOTA program.

In contrast to the abundant information available about Earth's impacts (most of them are detected visually or acoustically from the ground and from the space), only a limited amount of information about Moon impacts can be obtained solely from the detection of lunar flashes.

The kinetic energy $K$ of the impactor, can be estimated from the luminous energy $E_r$ of the flash (in a given spectral band), assuming a simple relationship:

\beq{eq:eta}
E_r = \eta K
\eeq

Here, $\eta$ is the so-called {\it luminous efficiency}.  This formula, though simple, \hll{have proved} to be very useful at reconstructing the intrinsic properties of impacts, using only the light emitted by the event. On Earth, for instance, the luminous efficiency of bright fireballs has been estimated to be of $\eta_{\rm fireballs} \lesssim 10\%$ \hl{\citep{Brown2002, Gritsevich2011, Bouquet2014}}.  Naturally, $\eta_{\rm fireballs}$ cannot be used to estimate the kinetic energy of lunar impacts (the physics of \hll{both processes} are very different). However, since on Earth we can estimate independently the velocity and initial mass of some meteors, we are confident that a simple relationship like Eq. (\ref{eq:eta}) can be used to relate $E_r$ and $K$.

In the case of impacts on the Moon, the estimation of the distribution of particle masses and diameters in major meteor showers (which depends on their kinetic energy and hence on $\eta$), has allowed us to constrain the value of this quantity within an interval $\eta\sim\etarange$, or equivalently $\log\eta\sim\logeta$ \citep{Madiedo2014}.

In general, $\eta$ will depend on the spectral region where the flash emission is measured.  The values we use here, have been found for observations in the visible spectrum (see eg. \citealt{Ortiz2006,Ortiz2015,Suggs2014}).

Kinetic energy alone does not allow the precise estimation of other physical properties of the impactor.  Speed, mass and composition are, for instance, almost impossible to be determined just from observation of the flashes.  However, ``educated guesses'' of the speed and incident angle (as obtained from independent theoretical models), provide useful estimations of the meteoroid properties.

The success of these theoretical models was subject to several tests in 2013. On march 17 a bright flash was spotted by researchers at NASA’s Marshall Space Flight Centre.  Later that year, LRO was able to spot the result of the impact, a fresh crater of 18.8 meters at the observed position of the flash.  The size of the crater was consistent with the theoretical predictions performed from the observed characteristics of the flash  \citep{Robinson2015}. The same year, a bright flash was spotted by MIDAS on September 11, 2013. \cite{Madiedo2014} estimated that, assuming a luminous efficiency $\eta=0.002$ and an impact speed of $\vimp=17$ km/s, the crater left by the impact has a diameter of 34 m.  The discovery by the LRO of a 46 m crater in the location of this flash, gives some confidence to the theoretical models.  Still, one or two cases are statistically insufficient to demonstrate, without any doubt, the validity of the models. The detection and analysis of more events will be required to test/verify existing models and to improve them.

\section{Data}
\label{sec:data}

Our analysis of the \LJ\ are based on several independent observations performed by amateur and professional astronomers around the world. Most of the observations used in this work were originally intended for different purposes that the ones we give them here (artistic purposes or citizen astronomy campaigns).

\begin{figure*}
    \centering
    \includegraphics[width=0.3\textwidth]{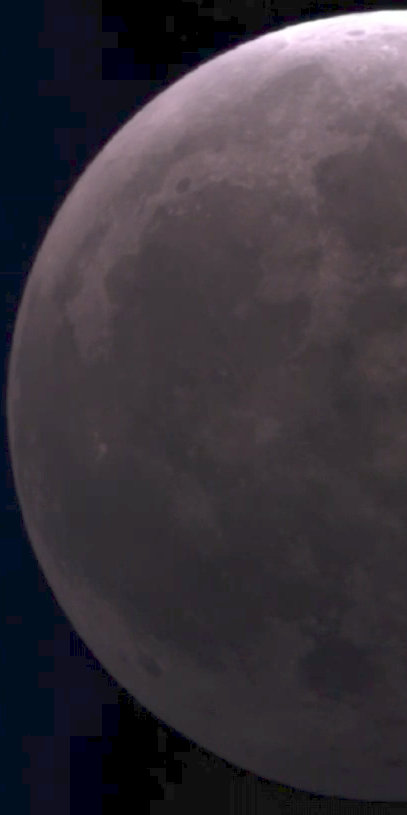}
    \includegraphics[width=0.3\textwidth]{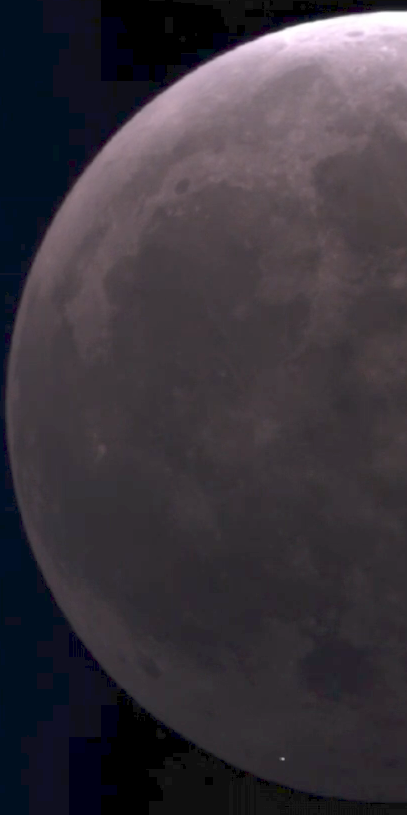}
    \includegraphics[width=0.3\textwidth]{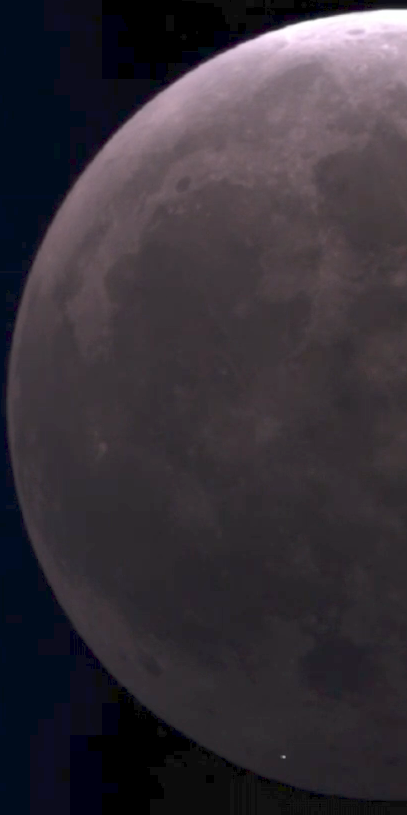}\\
    \includegraphics[width=0.3\textwidth]{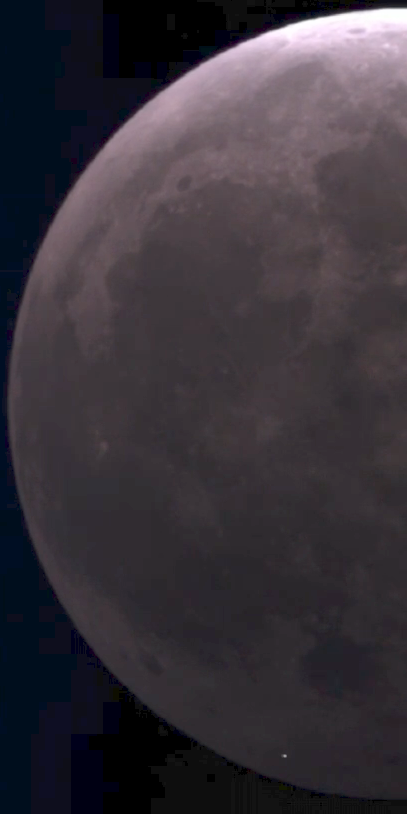}
    \includegraphics[width=0.3\textwidth]{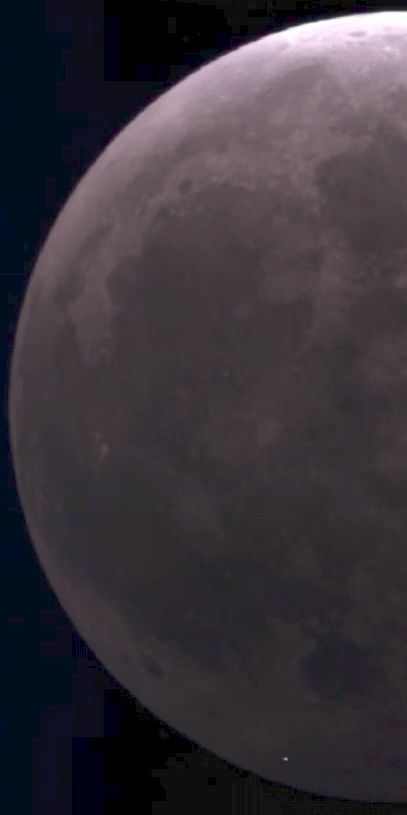}
    \includegraphics[width=0.3\textwidth]{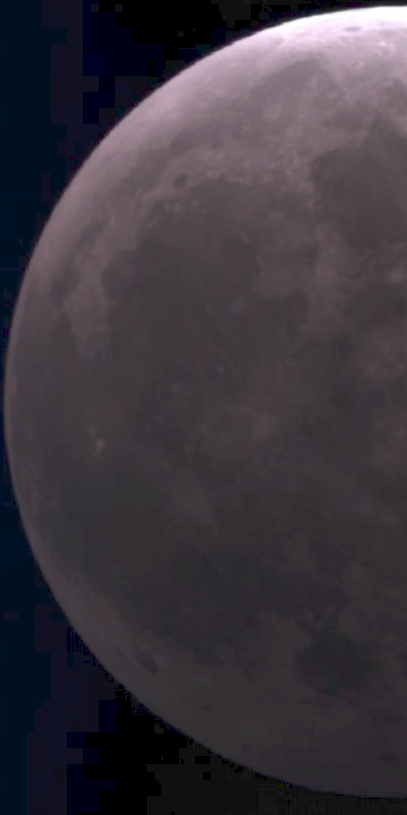}
    \caption{From left to right, frames of the video taken at Ouarzazate, Morocco by Time and Date mobile observatory.  The \LJ\ flash, which is visible close to the lower darker limb, appeared in four of the six frames.  Images reproduced with permission of Time and Date AS.}
    \label{fig:timeanddate}
\end{figure*}

The initial footage that motivated this research was a video taken by the mobile observatory of the \timeanddate\ astronomy portal\footnote{\url{https://www.timeanddate.com/}}. At the time of the eclipse, the observatory was located in Ouarzazate, Morocco. These observation supported the on-line transmission of the phenomenon in the \timeanddate\ website.
We extracted 6 frames \hl{from} the video, around the time of the flash and analysed them separately to obtain a light profile of the event and estimating the flash duration (see Section \ref{subsec:photometry}).  In Figure \ref{fig:timeanddate}, we show the selected frames.

Once the time of the impact was precisely determined, many casual observers around the world, looked-up among their images to see if the event was accidentally recorded.  Two of us (J.Z. and K.P.) received images and data from amateur astronomers in Colombia and the Dominican Republic. Independently, one of us (M.T.), received additional images via submission to the {\tt Astronomy Picture of the Day (APOD)}\footnote{\url{http://apod.nasa.gov}} and notice the potential for collective citizen science project, particularly in parallax information from simultaneous observation of the same events. The photographers were contacted and permission to use their images requested and subsequently granted. After reviewing the images, we selected 6 pictures meeting basic criteria of quality and metadata availability, which is required for their proper reduction.

In Table \ref{tab:locations} we present the properties of all the location from which we obtain pictures.


\begin{table*}
\centering
\resizebox{\textwidth}{!}{\begin{tabular}{cccccccc}
\hline\hline
Location & Latitude & Longitude & Elevation & Exposure time & Flash magnitude$^\dagger$ & Apparent Position$^\ddagger$ &  Selenographic Coordinates$^\ddagger$ \\
 & (deg) & (deg) & (m) & (seconds) & (G magnitude) & J2000 ($\alpha$, $\delta$) & (Lat., Lon.) \\\hline
Santo Domingo\\(The Dominican Republic) & 18.43567 & -69.96872 & 26 & 20.0 & $11.68\pm 0.46$ & (8.1826,20.2841) & (-29.67,-67.81)\\
Georgia\\(USA) & 32.51667 & -83.65440 & 107 & 16.0 & $10.76\pm 0.44$ & (8.1968,20.0198) & (-29.56,-67.84)\\
Boavista\\(Republic of Cape Verde) & 16.14361 & -22.86400 & 55 & 16.0 & $10.70\pm 0.47$ & (8.1297,20.2223) & (-29.51,-67.89)\\
Santa Cruz de Tenerife\\(Cannary Island, Spain) & 28.14169 & -16.62200 & 1187 & 2.0 & $8.52\pm 0.43$ & (8.1291,19.9900) & (-29.43,-67.84)\\
Karben\\(Germany) & 50.21615 & 8.79607 & 140 & 1.0 & $8.54\pm 0.49$ & (8.1342,19.5625) & (-29.29,-67.97)\\
Velky Osek\\(Czech Republic) & 50.09820 & 15.18885 & 192 & 0.5 & $6.61\pm 0.47$ & (8.1334,19.5386) & (-29.40,-67.94)\\
Vienna\\(Austria) & 48.25000 & 16.21700 & 450 & 4.0 & $9.88\pm 0.44$ & (8.1315,19.5562) & (-29.06,-67.99)\\

\hline\hline
\multicolumn{8}{l}{$^\dagger$ Section \ref{sec:energy}, $^\ddagger$ Section \ref{sec:location}.}
\end{tabular}}
\caption{\hl{Properties of the images analyzed in this work.}\label{tab:locations}}
\end{table*}

\begin{figure*}
    \centering
    \includegraphics[width=0.9\textwidth]{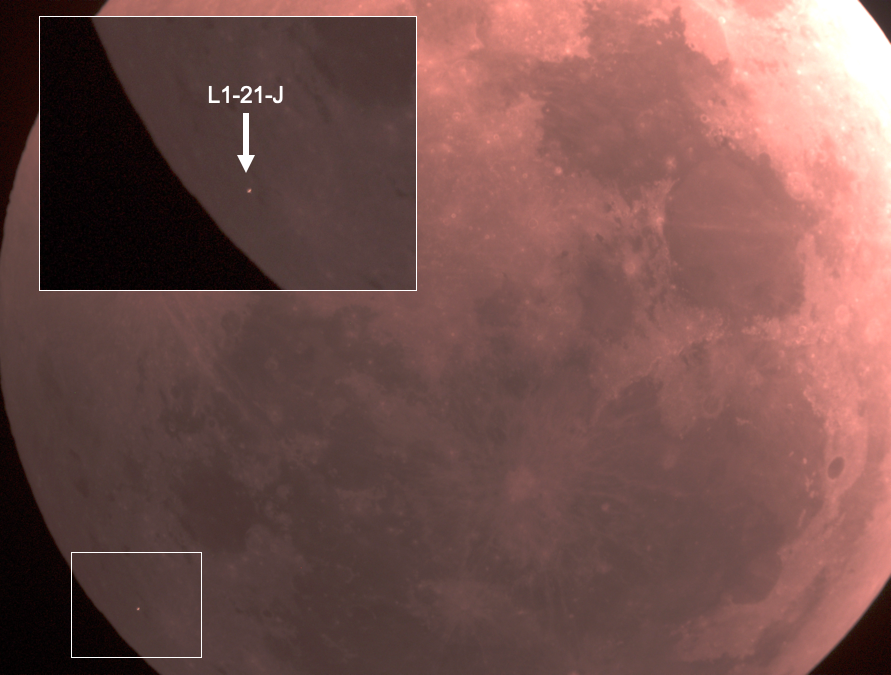}
    \caption{Picture of the total lunar eclipse at the time of the impact flash taken the observatory \LaLoma, Colombia. Picture by Jonathan Ospina, Mauricio Gaviria and Sergio L\'{o}pez.}
    \label{fig:laloma}
\end{figure*}

The picture with the largest resolution (that was used for locating the impact site, see Section \ref{sec:location}), is a short exposure (0.71 seconds) taken with the 25" (635 mm) FL = 2700 mm telescope of \LaLoma\ Observatory in San Vicente Ferrer (Antioquia) in Colombia.  The picture (see Figure \ref{fig:laloma}) was taken using a 17.7$\times$13.4 mm (4656$\times$3520 pixels) CMOS ZWO ASI1600MC detector, working with a f/2.92 focal reducer in the primary focus of the telescope, yielding an effective FL = 1854.2 mm.  Given the large aperture and low f-value of the telescope, the event was captured with a relatively short exposure.  The picture was taken at 04:41:37 UTC, which also coincide with the time reported by MIDAS.

Each pixel of the camera attached to \LaLoma\ observatory telescope, covers 0.32 arc sec, that in ideal atmospheric conditions correspond to a spatial resolution of 0.7 km/px
on the surface of the Moon at the centre of its face.
At the location of the impact ($\sim 60$ degrees from the centre of the near side), the resolution will be larger than 2 km/px (again, under idealised atmospheric conditions). However, since the actual seeing at the time of the picture was a few arc sec, the actual resolution downgraded to $\gtrsim 20$ km/px.

\begin{figure*}
    \centering
    \includegraphics[width=0.8\textwidth]{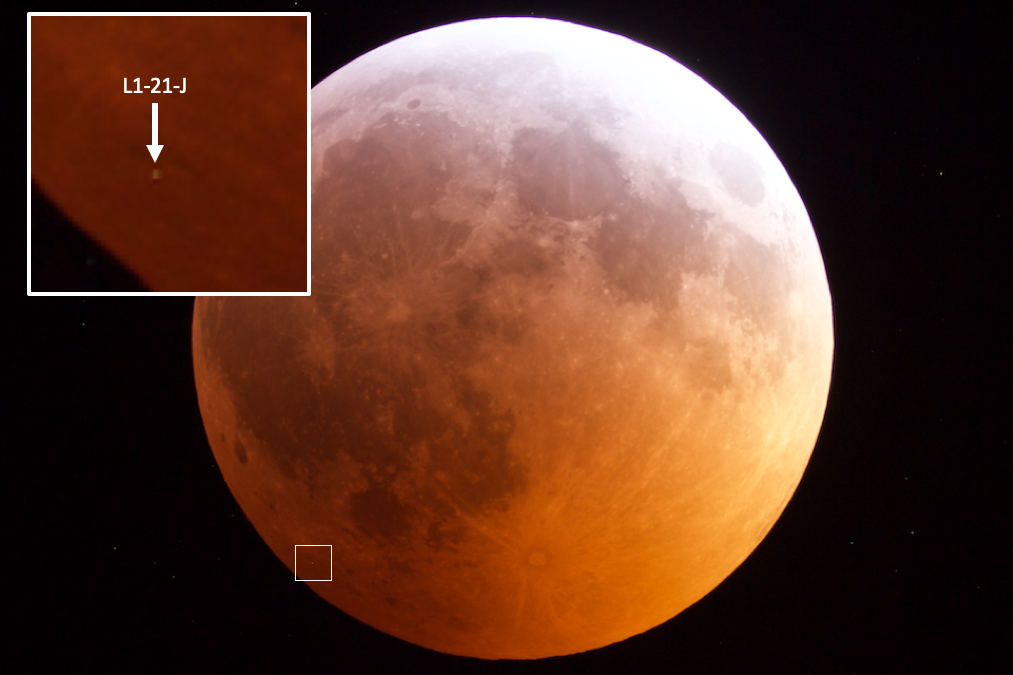}
    \includegraphics[width=0.8\textwidth]{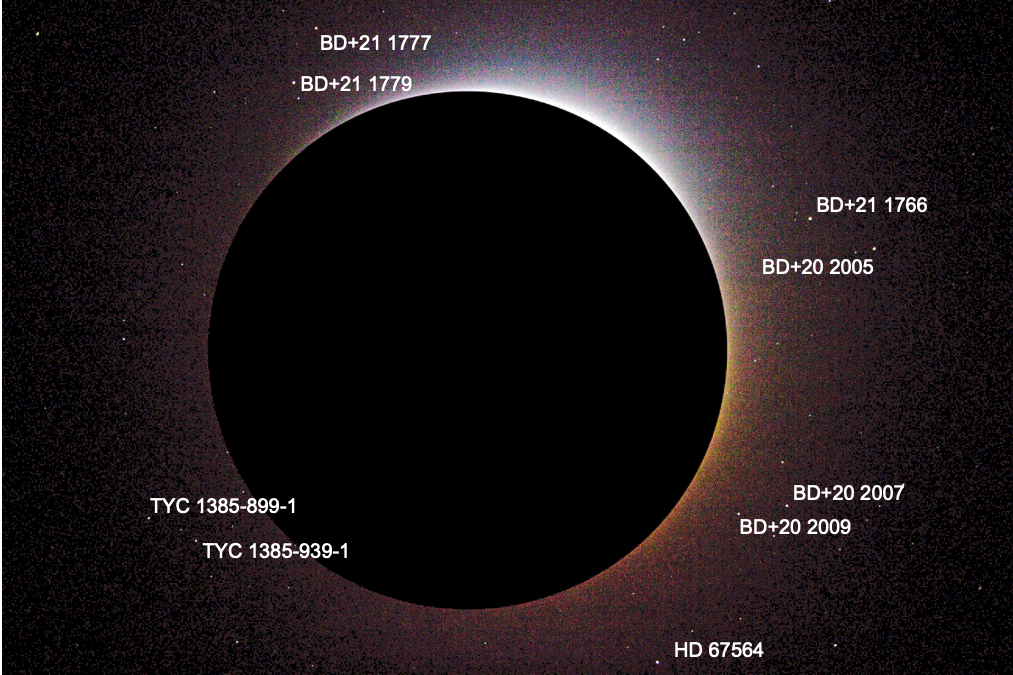}
    \caption{Picture of the total lunar eclipse at the time of the impact flash taken in Santo Domingo, the Dominican Republic.  Upper panel: detail of the \LJ\ flash.  Lower panel: the same picture with the moon removed, highlighting the background stars.  9 stars were identified and used for the parallax and photometry analyses.  Picture by Fritz Pichardo.}
    \label{fig:fritz}
\end{figure*}

The remaining, lower resolution pictures, contained background stars and they were used for a parallax-based (independent) estimation of the impact site (Section \ref{sec:parallax}) and for the photometry of the flash (Section \ref{sec:energy}). For illustration purposes, we show In Figure \ref{fig:fritz}, the picture taken by Fritz Pichardo in Santo Domingo, the Dominican Republic.  This 20-seconds exposure, started at 04:41:24 UTC and lasted until 04:41:44 UTC enclosing the time of the flash.  The picture was taken using a Canon T3i DSLR camera (18Mpx APC-S CMOS sensor), installed on the secondary focus of an equatorial mounted 8 inch Celestron CPC 800 Schmidt-Cassegrain telescope, with a focal length FL=2032 mm (f/10).  The camera was installed with a focal reducer f/6.3, yielding an effective FL=1280.16 mm.

The other 6 pictures had the following characteristics. A 16 seconds exposure taken by Petr Hor\'{a}lek in Boa Vista, Cape Verde. The image was taken using a Canon EOS 6D DSLR camera attached to a MTO 1100/f10.5. Gregory Hogan took a 16 seconds exposure from Kathleen, Georgia, USA using a Canon EOS 6D DSLR camera. Fritz Helmut Hemmerich captured the event with 2 seconds exposure time from Tenerife in the Canary Islands using a RASA 11"/F2.2, ASI 071 colour camera (cooled to -25$^{\circ}$C, at the highest dynamic range). From Velk\'{y} Osek, Czech Republic, Libor Haspl took a 0.5 seconds exposure using a Canon 5D Mk IV mounted on a 8" telescope. From Vienna, Austria, Robert Eder Artis took 4 seconds exposure from his Canon 600Da astro modified DSLR camera with Skywatcher Newton 130/650 PDS. Lastly, we received from Dr. Sighard Schr\"abler and Dr. Ulrike L\"offler from Karben, Germany a 1-second exposure taken with a Sony A7s DSLR camera mounted at the primary focus of 12" Foto-Newton reflector.

For reproducible purposes, we provide access to all the footage used in this work (raw images) in a companion {\tt GitHub} repository\footnote{\url{https://github.com/seap-udea/MoonFlashes}}.


\section{Location}
\label{sec:location}

Determining the precise location of the impact from images that were not properly calibrated for this purpose, is challenging.  Here, we devise two independent procedures: a visual comparison between the highest resolution picture and LRO maps, and a parallax-based location estimation (geometrical procedure).

\subsection{Visual procedure}
\label{subsec:vislocation}

In Figure \ref{fig:location} we graphically summarise our visual procedure. We first superimpose and align our highest resolution picture (see Figure \ref{fig:laloma}) with Lunar Reconnaissance Orbiter Camera (LROC) orthographic projection maps\footnote{\url{https://quickmap.lroc.asu.edu}} of the southeast lunar quadrant.
Then, we distort the flash image and superimpose it to a cylindrical equidistant projection of the selected region (lowest panel in Figure \ref{fig:location}). This superposition allowed us to estimate the coordinates of the impact site and their corresponding errors.

According to our analysis, the impact happened to the southeast of {\it Mare Humorum}, near to the easily identifiable {\it Byrgius} crater and inside a triangle with vertices in the {\it Lagrange} H, K and X craters. The impact signature (crater and/or rays) should be inside an almost elliptical region centred at lat. \latvisual, long. \lonvisual\ with an east-west 18-km major axis, and a north-south 15 km minor axis. This is a reasonable-sized area, where prospecting satellites may look for a crater in the near future (see Section \ref{sec:sizes}).

\subsection{Geometrical procedure}
\label{subsec:geolocation}

In all pictures, we achieved to identify reference stars (see for instance the lower panel of Figure \ref{fig:fritz}). The position and brightness of these stars provide us valuable information for performing the astrometry and photometry on the images. In Tables \ref{tab:referencestars1} and \ref{tab:referencestars2}, we provide detailed information of all the stars identified in our images, including their sky coordinates, namely J2000 (\RA,\DEC), image (centroid) coordinates, $(X,Y)$, magnitudes and counts on each image channel (see Section \ref{sec:energy}).


\begin{table*}
\centering
\begin{tabular}{cccc}
\hline\hline
Surface Feature & Selenographic Coordinates & Geocentric Position$^\dagger$ & Distance$^\dagger$ \\
                & (Lat.,Lon.)               & J2000 ($\alpha$, $\delta$)      & (km)     \\\hline
Moon Center & - & (8.1667,20.4362) & 357046\\
L1-21J & (-29.42,-67.90)$^\ddagger$ & (8.1799,20.2505) & 356553\\
Byrgius A & (-24.54,-63.83) & (8.1804,20.2709) & 356363\\
Grimaldi & (-5.53,-68.26) & (8.1840,20.3505) & 356135\\
Aristachus & (23.69,-47.49) & (8.1815,20.5024) & 355491\\
Plato & (51.64,-9.30) & (8.1721,20.6433) & 355636\\
Tycho & (-43.40,-11.26) & (8.1659,20.2433) & 356271\\
Copernicus & (9.64,-20.06) & (8.1737,20.4634) & 354999\\
Manilius & (14.44,9.06) & (8.1645,20.5177) & 355303\\
Dionysus & (2.77,17.29) & (8.1608,20.4727) & 355576\\
Chladni & (3.47,-0.23) & (8.1662,20.4603) & 355284\\
Kepler & (8.15,-37.99) & (8.1789,20.4370) & 355312\\
Bullialdus & (-20.75,-22.30) & (8.1715,20.3201) & 355721\\

\hline\hline
\multicolumn{4}{l}{\footnotesize $^\dagger$ Calculated geocentric coordinates and distance (see Section \ref{sec:parallax})}\\
\multicolumn{4}{l}{\footnotesize $^\ddagger$ Calculated with our geometrical procedure (see Section \ref{subsec:geolocation})}\\
\end{tabular}
\caption{Lunar features reference points and their selenographic coordinates, along with the apparent geocentric equatorial coordinate RA, Dec and geocentric distance results from parallax analysis.}
\label{tab:surfacefeatures}
\end{table*}

\begin{figure*}
    \centering
    \includegraphics[width=0.7\textwidth]{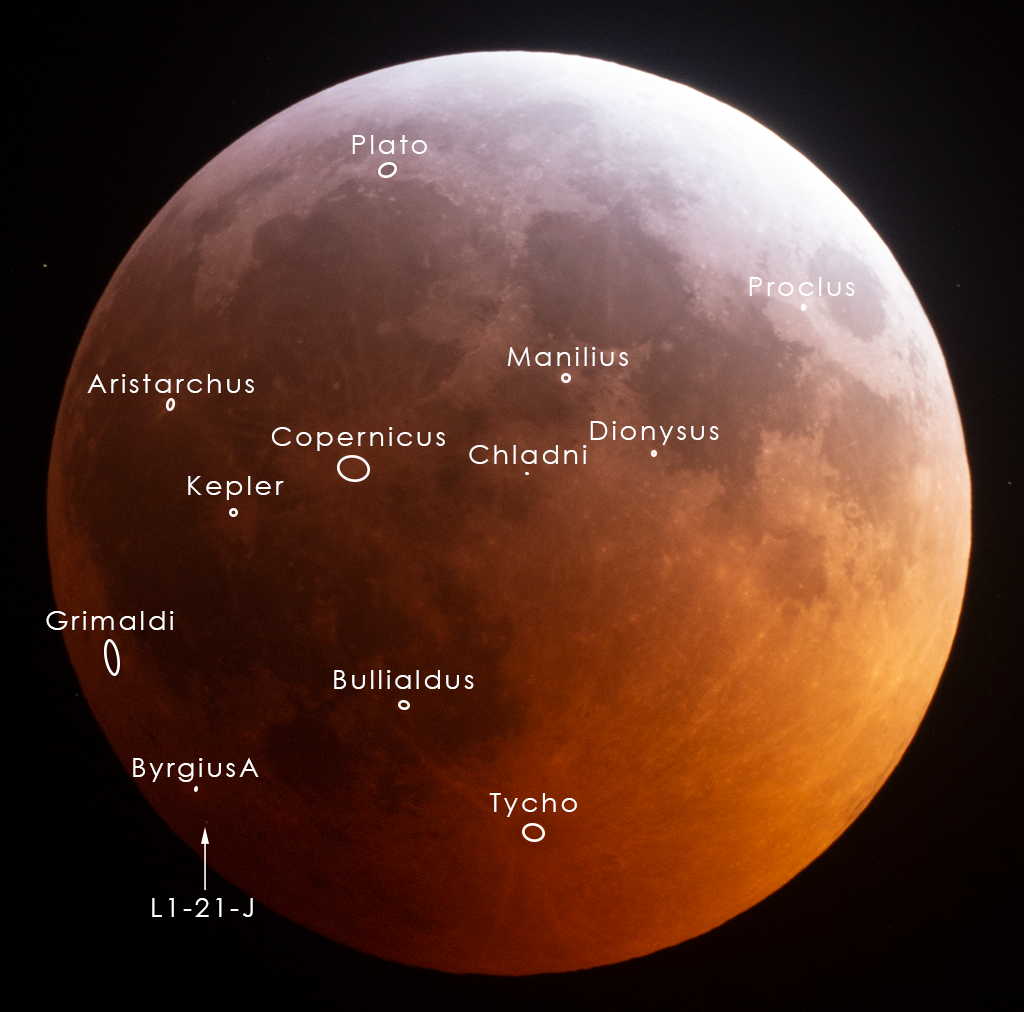}
    \caption{Craters and lunar features used by the Geometrical procedure in calibrating for the selenographic coordinates (section \ref{subsec:geolocation}).  Picture by Petr Hor\'alek.}
    \label{fig:petr}
\end{figure*}

Similarly, we identified and measure the position over all the images of 11 surface features (mainly large craters) as shown in Figure \ref{fig:petr}.  In Table \ref{tab:surfacefeatures} we provide the selenographic latitude and longitude of these features, along with their calculated geocentric sky position and distance, as estimated with the procedure below.

Since our aim here is to estimate the selenographic location of the impact, we need a method to convert from image to selenographic coordinates.  We perform this transformation in two steps. First, we calculate, for each image, the so-called {\it plate constants}, namely the coefficients of a linear (affine) transformation that convert image into sky coordinates and vice versa.  The resulting projected sky position of the impact site, as computed with this procedure, are provided in Tables \ref{tab:referencestars1} and \ref{tab:referencestars2}.  We apply this transformation to compute also the projected position in the sky of the selected surface features.

Transforming selenographic into sky coordinates, involves a complex rotation in the sky and the precise knowledge of the relative Earth-Moon position.  We can also model this transformation, by a general formula:

\begin{eqnarray}
\label{eq:selenographictoradec1}
\alpha &=& a_1\sin\left(\lambda-\lambda_0\right)\cos\left(\phi-\phi_0\right)\nonumber\\
&& +a_2\sin\left(\phi-\phi_0\right)+a_3\\
\label{eq:selenographictoradec2}
\delta &=& a_4\sin\left(\lambda-\lambda_0\right)\cos\left(\phi-\phi_0\right)\nonumber\\
&& +a_5\sin\left(\phi-\phi_0\right)+a_6,
\end{eqnarray}

where $a_1,...,a_6$ are free coefficients and $(\phi_0,\lambda_0)$ are the (unknown) selenographic coordinates of the centre of the moon (determined by lunar libration).  Since we know the values of $\alpha$, $\delta$, $\lambda$ and $\phi$ of at least 11 points on the surface (the lunar features in Table \ref{tab:surfacefeatures}), we can find the best-fit values of the 8 free parameters of this general transformation. Once we have the parameters of the selenographic to sky coordinates transformation, we may invert it to estimate the position of the impact.  The resulting sky and lunar coordinates obtained for each image with this procedure are reported in the last two columns of Table \ref{tab:locations}.

\begin{figure*}
    \centering
    \includegraphics[width=0.45\textwidth]{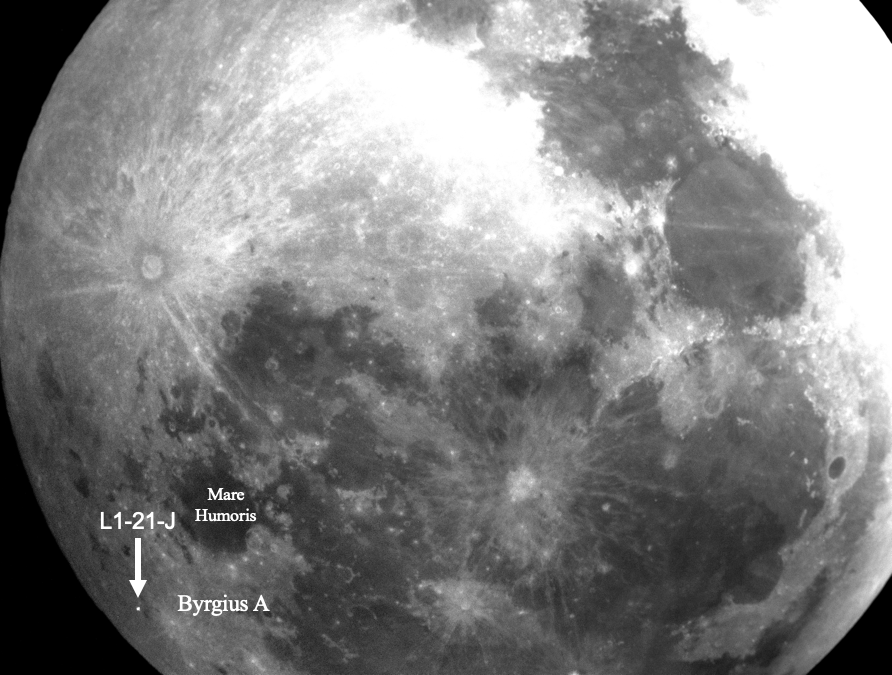}
    \includegraphics[width=0.45\textwidth]{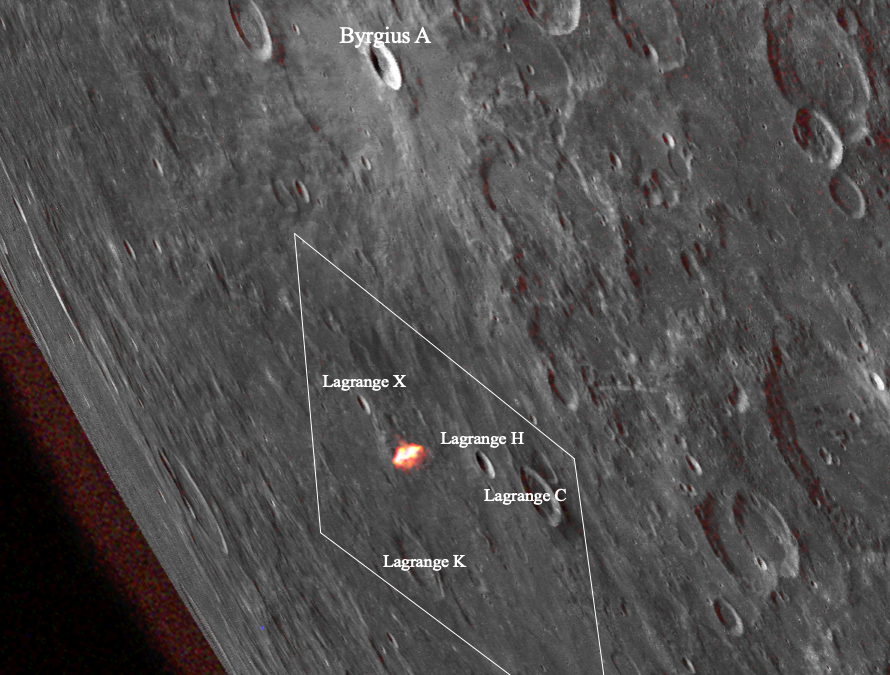}\\
    \includegraphics[width=0.45\textwidth]{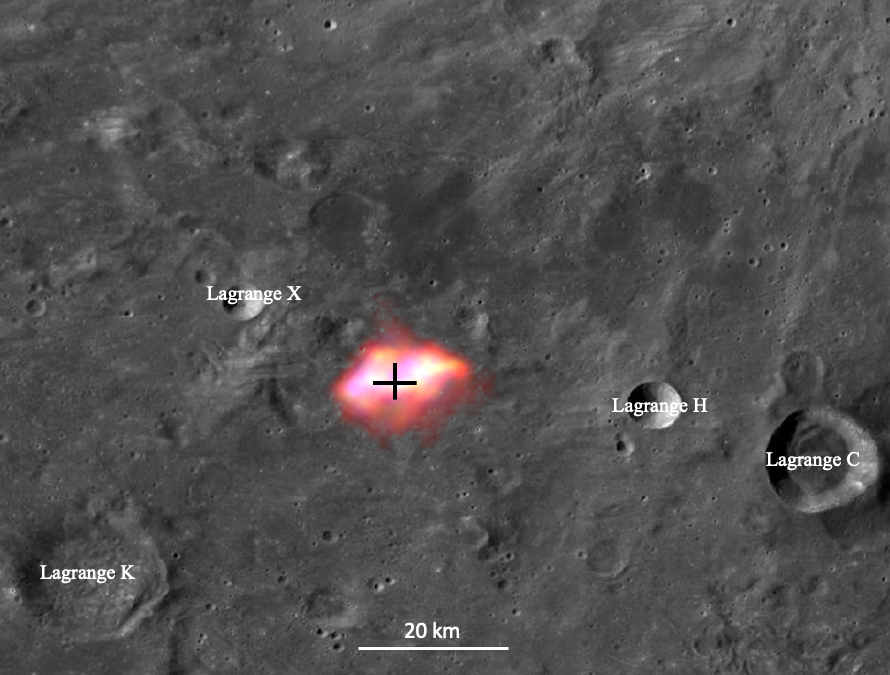}
    \includegraphics[width=0.45\textwidth]{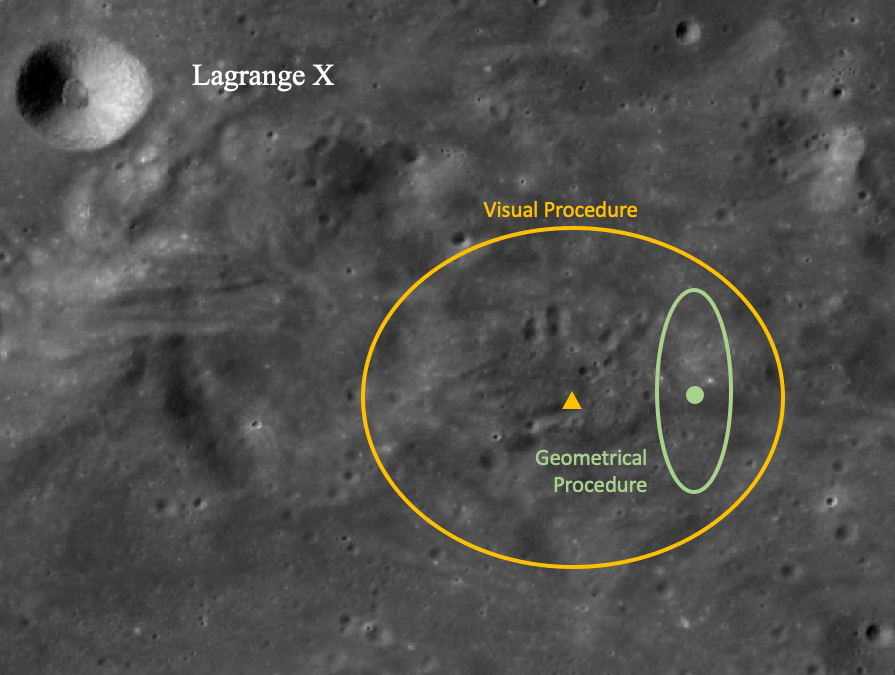}
    \caption{Impact estimated location. Upper row: original picture taken at \LaLoma\ observatory in Colombia (left) and superposition of the original flash image and a LROC ortographic map (right). Bottom row: image of the impact flash superimposed to a high resolution LROC cylindrical map (left), including error ellipses associated to the visual and geometrical procedures (right).}
    \label{fig:location}
\end{figure*}

We predict that the impact happened inside \hll{an ellipse} with lat. $\latgeo$ and lon. $\longeo$.  This result is in fair agreement with our visual estimation (see Figure \ref{fig:location}).

\section{Orbit}
\label{sec:orbit}

We cannot reconstruct the orbit of an object impacting the Moon using only the observation of its associated light flash. Still, and as we will demonstrate in this section, we can use the time and location of the impact to theoretically constraint the incoming direction and speed of the object.

For this purpose, we apply {\it Gravitational Ray Tracing} (GRT), a novel numerical technique recently introduced by \cite{Zuluaga2018} with the aim of computing efficiently, meteoroid impact probabilities on the surface of any planetary body in the Solar System.  The method was tested on Earth and it successfully reproduced the impact speed distribution of moderately large fireballs \citep{Zuluaga2018}.

In GRT, random incoming directions (elevations and azimuths) are generated following a blue-noise distribution on the sky above the impact site (see \citealt{Zuluaga2018} for a detailed explanation). These random directions are combined with a set of regularly spaced impact speeds, in the interval between the Moon escape velocity and the escape velocity from the Solar System at the distance of the Earth-Moon system, to build many different random initial conditions.  For each initial condition, the trajectory of a test particle is integrated backwards in time in the gravitational field of the Solar System. After one year, the heliocentric orbital elements of the test particle (namely its asymptotic orbit) are compared against the orbital elements of a target population (NEOs, sporadic meteoroids, a mono-kinetic population, etc.)

The relative probability that the actual meteoroid comes from one of the many directions and impact speeds in the simulation, is proportional to the number density of objects in the target population, in the space of classical orbital elements, as computed around the asymptotic orbit.  In other words, a given initial condition is more probable if many potential parent objects in the target population have orbital elements similar to that of the asymptotic orbit associated to that condition. Further details of the technique are found on the original paper by \cite{Zuluaga2018}.

For our purpose here, we generate 997 random incoming directions on the sky above the impact location, with a minimum separation of 5$^\circ$ (incoming directions are not random, but carefully arranged to have a minimum separation with its closest neighbours; this configuration is intended to avoid numerical artefacts arising from the sampling procedure).  We also choose 100 regularly spaced impact speeds, that together with the incoming directions, create a set of 49901 different initial conditions.

41167 test particles (82$\%$) survived the numerical experiment, meaning that they did not collide against the Moon, the Earth or the Sun, nor escaped from the Solar System after being perturbed by the planets.

Using the relative probability of each test particle, we can now estimate the {\it marginal probability distributions} (mpd) of any dynamical or spatial impact property.  Thus, for instance, the probability that the impact speed is between $v$ and $v+\Delta v$ is simply the sum of the (relative) probabilities of all initial conditions having incoming speeds in this interval. The same method can be used to compute the mpd of the impact angle or radiant position.

The results of GRT \hl{depend} on selecting the right target population for the parent body. For relatively small impacts on the moon, it is customary to assume that most of the objects come from the population of small meteoroids and dust particles around the Earth-Moon system \citep{Halliday1996, Brown2002, Campbell2007, Madiedo2015a, Avdellidou2019}.  These particles are mostly cometary in origin \citep{Williams2014,Jenniskens2017} and have large velocities with respect to the Earth and the Moon \citep{Halliday1996, Campbell2007, Wiegert2009}. On Earth, those particles are mainly responsible for the so-called {\it sporadic meteor background} \citep{Campbell2007, Wiegert2009}.

\begin{figure}
    \centering
    \includegraphics[width=0.5\textwidth]{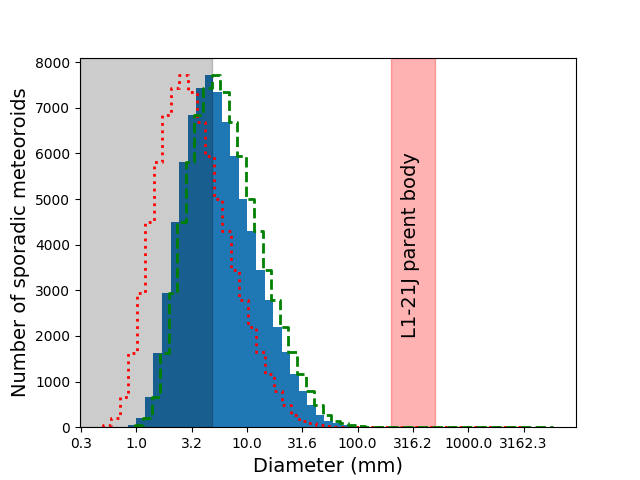}
    \caption{Distribution of sizes of the particles involved in the sporadic background.  \hl{For computing the meteoroid sizes we have assumed a common average density of 1000 kg m$^{-3}$.  In order to show the effect that assuming different densities have in the distribution, we also show (empty histograms) the corresponding distribution assuming extreme average densities of 500 kg m$^{-3}$ (green dashed) and \hll{3700} kg m$^{-3}$ (red dotted)}.  The Gray shaded area \hl{corresponds} to sizes below the detectable thresholds.  The red shaded area \hl{shows} the range of sizes estimated for the L1-21J impactor.}
    \label{fig:sporadics}
\end{figure}

In Figure \ref{fig:sporadics} we show the distribution of particle diameters of more than 80,000 sporadic meteors captured between 2010 and 2013 by the CAMS survey \citep{Jenniskens2016a,Jenniskens2016b}\footnote{\url{http://cams.seti.org/}}.  Diameters have been calculated using the integrated brightness and impact velocity provided by the CAMS database, and the methods and formulas in \cite{Jacchia1967}. \hl{For the nominal case,} we have assumed an average particle density $\rho=1000$ kg m$^{-3}$. The CAMS database is complete for particles larger than $\sim$5-34 mm \citep{Jenniskens2016b}.  For comparison purposes, we included the estimated size of the parent body of L1-21J, as estimated by \cite{Madiedo2019} and this work (see Section \ref{sec:sizes}).

\begin{figure}
    \centering
    \includegraphics[width=0.45\textwidth]{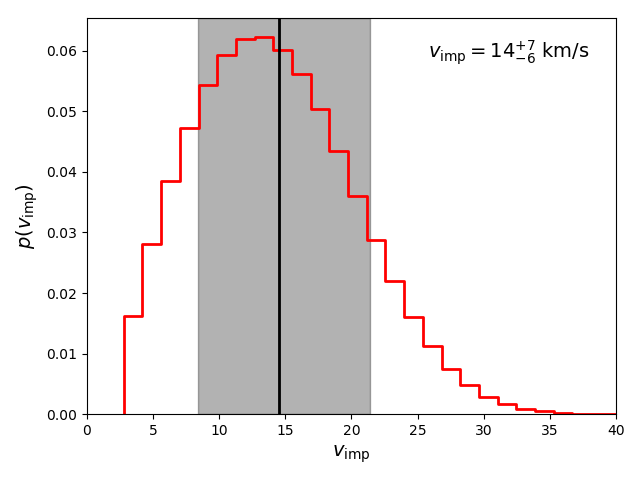}
    \includegraphics[width=0.45\textwidth]{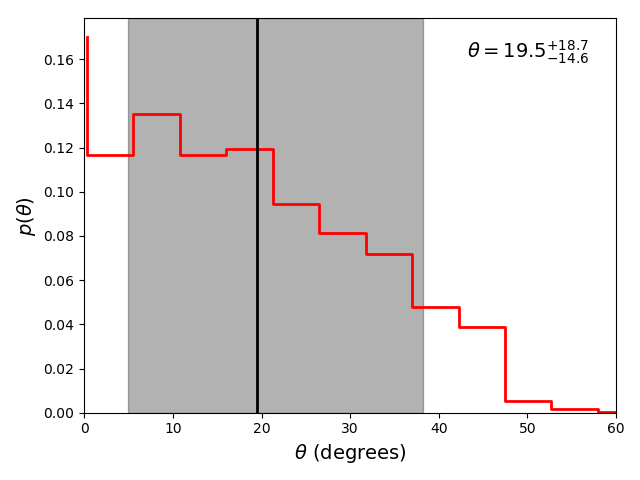}
    \caption{Marginal probability distributions (mpd) of Impact speed (upper panel) and incident angle (lower panel), as computed using a GRT analysis at the time and location of \LJ\ event.}
    \label{fig:ppds}
\end{figure}

\hl{A simple inspection of the CAMS survey data reveals that} only \hl{0.03-0.09\% (depending on the assumed densities)} of the objects in the sporadic background have diameters larger than 10 cm \hl{(in a size range where the sample is considered complete)}. 

If we assume that most of the objects producing lunar impacts come from this population, and fall into the moon to a rate of $\sim$0.4 impacts per hour \citep{Halliday1996, Suggs2014}, the probability that one sporadic background object larger than 10 cm impacted the moon during a total lunar eclipse is \hl{0.01--0.03}\%.  This would mean that, in average, we would must wait $\sim$\hl{3 000--10 000} total lunar eclipses, to witness an impact as brilliant as L1-21J caused by \hll{a sporadic} background object. 

\hl{The most natural alternative to consider the parent body of L1-21J a sporadic meteor (cometary origin), is assuming that it belonged to the population of Near Earth Objects (NEOs), most of which are asteroidal in nature \citep{Borovivcka2015}.  For the purpose of GRT, this assumption requires to know the orbital distribution of NEOs at the size range of the L1-21J impactor. Regretfully,} the distribution of NEOs having diameters below $\sim$50 m is still unknown (see eg. \citealt{Granvik2017}). \hl{Therefore,} if the object impacting the Moon on January was \hl{for instance} a small fragment of \hl{a larger NEA}, it probably had a very different orbit than that of its parent asteroid (see eg. \citealt{Babadzhanov2012}).

\hl{Still, and in the absence of better information, we can assume that the distribution of small NEOs is not too different than that of larger ones. Although this hypothesis is relatively weak, from a rational point of view ({\it Occam's razor} principle) it is much better than assuming that we witnessed the impact of an extremely rare object (one, having a large relative velocity, ie. belonging to the sporadic background).}

In Figure \ref{fig:ppds} we show the ppds of the impact speed and incident angle (elevation) at the time and location of \LJ\ event as calculated with GRT assuming that the parent body population was that of the already discovered NEOs. As expected, the impact speed obtained with GRT, $\vimp=\vimpvalue$ km/s is lower than the typical values assumed for this quantity in the context of lunar impact studies, namely $16-22$ km/s, where sporadic background objects or mono-kinetic sources, are assumed as the most probably parent population \citep{Ivanov2006,Madiedo2014,Suggs2014}. Still, since our method does not assume a single value for this quantity but instead uses the whole speed distribution, our predictions will surely overlap that obtained by other authors.


\begin{figure*}
    \centering
    \includegraphics[width=0.48\textwidth]{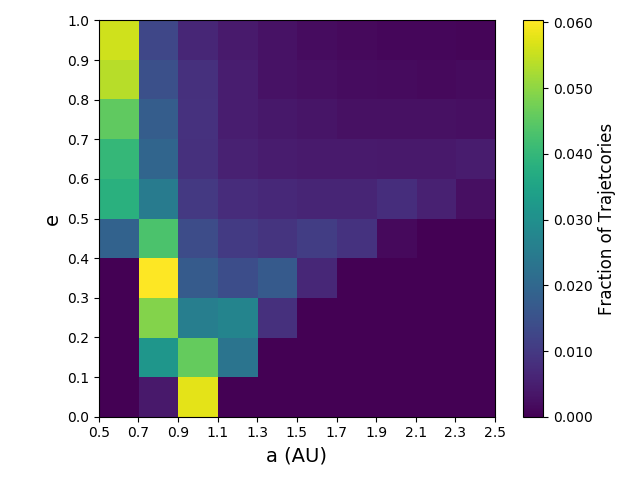}
    \includegraphics[width=0.48\textwidth]{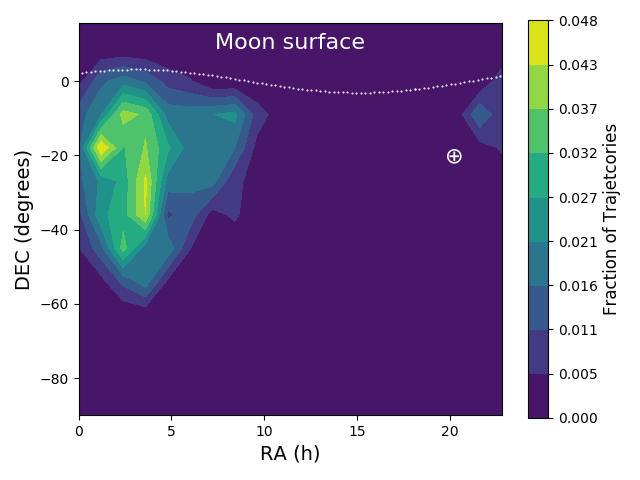}
    \caption{Heat maps of the two-dimensional marginal probability distribution of orbital elements of the parent body (left panel) and radiant locations on the sky (right panel).  The dotted line in the right panel represent the boundary between the sky and the solid moon at the impact site. The Sun (and the Earth during the total lunar eclipse) are located in the sky above the impact site close to the $\oplus$ symbol.}
    \label{fig:maps}
\end{figure*}

Finally, we calculate the two-dimensional marginal probability distribution of the orbital elements of the parent body and its radiant on the sky above the impact site.  The results are shown in Figure \ref{fig:maps}.

The distribution of orbital elements seems to favour the hypothesis that the impactor came from an orbit inside the Earth's orbit.  If asteroidal in origin, its parent body could be probably classified as an Aten.

Regarding its incoming direction in the sky, our simulation seems to favour an extensive region around RA$\sim\RArad$ h, DEC$\sim\DECrad^\circ$.  No major meteor shower has a radiant around those positions.

\section{Energy of the impact}
\label{sec:energy}

Flashes are the result of thermal emission from hot plasma plumes created by vaporised material coming from the meteoroid and the surface \citep{Rubio2000}. The impact event takes place in a very short time, ie. $\sim 10^{-2}$ s, but visible light-emission last within 0.05-0.1 s. In order to estimate the energy of the impact it is necessary to measure the brightness and duration of the flash.

\subsection{Photometry}
\label{subsec:photometry}

To illustrate the method we use for estimating the average flash brightness, we use as a reference image, the picture taken by Fritz Pichardo in the Dominican Republic (see Figure \ref{fig:fritz}).  In the picture, the flash was detected at a ${\rm SNR}\sim 10$.  At least 9 well-known stars were also identified with similar SNR.

In Table \ref{tab:referencestars1} and \ref{tab:referencestars2} we show the properties of the reference stars, along with the value of the counts detected around their position (as determined with aperture photometry, AP) in each image channel (RGB).  For completeness, we also include the counts detected on the flash position.

Performing precision photometry with RGB images is challenging. Although the spectral response of the camera sensor is well-known \citep{Deglint2016} and it covers all the visible spectra in a similar fashion that standard photometric filters, understanding how to relate the counts in each channel to a magnitude in a specific photometric system is not trivial.

We assume for simplicity that the Gaia G magnitude (which is already known for all the reference stars and also covers a wide region of the visible spectrum) could be in principle calculated from the counts in the RGB channels $C_R$, $C_G$ and $C_B$, using the formula:

\beq{eq:photometry}
G = Z + k_R \log C_R + k_G \log C_G + k_B\log C_B
\eeq

Here $Z$ is the unknown zero-point and $k_R$, $k_G$ and $k_B$ are also unknown constants.

The value of these constants were found by fitting with the previous formula the G magnitude of the reference stars on each image.

In Table \ref{tab:locations} we present the G magnitude of the flash as measured after performing the AP photometry of our seven images, along with their corresponding errors.

Since exposure time for the reference stars in each image is larger than the duration of the flash ($t_f=\tflash$ s), The magnitude $G(t)$ (with the exposure-time $t$) estimated with this procedure will under- or over-estimate the actual average magnitude $G_f$ of the flash. $G(t)$ and $G_f$ are related by the Pogson's law:

\beq{eq:exposureline}
G(t)=G_f-2.5\log(t/t_f)
\eeq

\begin{figure*}
    \centering
    \includegraphics[width=0.45\textwidth]{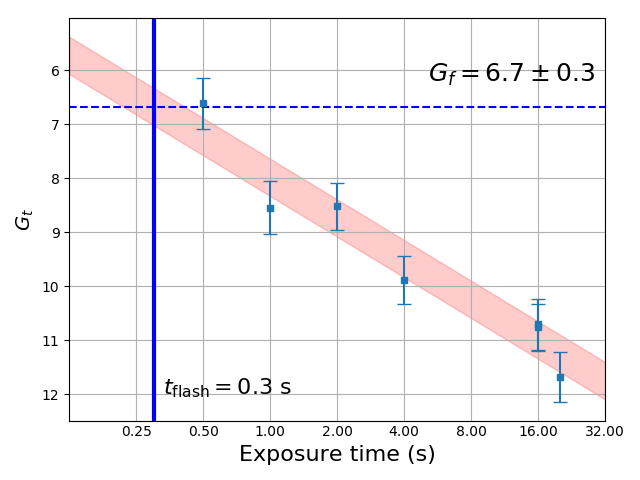}
    \includegraphics[width=0.45\textwidth]{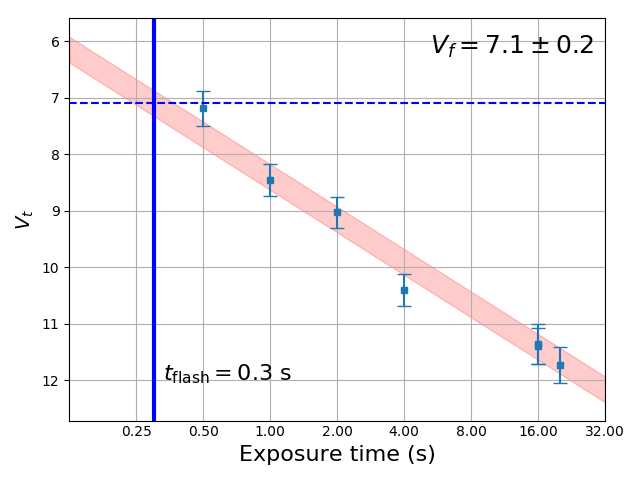}
    \caption{Average $G(t)$ and $V(t)$ magnitudes as estimated from each image having different exposure times. The intercept at the flash duration $t=0.3$ s provides an estimation of the true flash magnitudes, $G_f$, $V_f$.}
    \label{fig:Gmag-time}
\end{figure*}

Therefore, if we have independent $G(t)$ values, the true-average magnitude of the flash $G_f$ can be estimated by finding the intercept of the best-fit line in Eq. \ref{eq:exposureline} with the $\log t_f$ vertical line.  In Figure \ref{fig:Gmag-time} we show the result of applying this procedure to our observations of the \LJ\ event.

We find that combining all this data the flash magnitude is:

\beq{eq:Gf}
G_f=\Gf
\eeq

where the error arises from the dispersion of the G magnitudes estimated for different images.

We attempted to measure the relative brightness profile of the flash using the {\tt timeandddate} video. For that purpose we performed aperture photometry on the flash image on each frame.  We find that, for the sensitivity and time resolution of the video camera, no significant variation in the brightness was detected during the $\tflash$ seconds of the flash.

We repeated the previous procedure to find, for comparison purposes, the average V magnitude of the flash.  We find  $V_f=7.1\pm 0.2$ (see right panel in Figure \ref{fig:Gmag-time}).

\subsection{Luminous energy}
\label{subsec:luminous}

From the estimated magnitude in the G filter, we can estimate the visible luminous energy of the impact \citep{Madiedo2015b}:

\beq{eq:Er}
E_r = b_G 10^{(-G_f+G_0)/2.5} \Delta t\Delta\lambda_G f\pi R^2
\eeq

where $R=\Dimpgeo$ km is the geocentric distance to the flash at the time (calculated using the procedure in the Appendix), $b_G=\bG$ W/m$^2$/nm, $G_0=\Go$ and $\Delta\lambda_G=\deltaLambda$ nm are the calibration properties of the filter\footnote{Filter profile service \url{http://svo2.cab.inta-csic.es/svo/theory/fps3/index.php}}, $\Delta t$ is the flash duration, and $f$ is the degree of anisotropy of the light emission (if the light was emitted isotropically from the surface then $f=2$; conversely, if light is emitted at a very high altitude then $f=4$, see eg. \citealt{Madiedo2015b}).

Using the estimated (average) flash magnitude $G_f=\Gf$ and assuming an intermediate value $f=\fany$, we may finally estimate the total luminous energy released during the flash:

\beq{eq:logEr}
\log \left(\frac{E_r}{{\rm J}}\right)=\logEr
\eeq

\hl{Other authors (see eg. \citealt{Suggs2014,Bonanos2018,Madiedo2014,Madiedo2019, Yanagisawa2002}) use different approaches to estimate $E_r$.  If you have a light profile $I(t)$ (with $I$ the {\em luminous rate} measured in J s$^{-1}$), obtained with a camera with an exposure-time per-frame $\delta t$, you may calculate two quantities: 1) the peak luminosity $E_{\rm peak}=I_{\rm max} \delta t$ \citep{Suggs2014,Bonanos2018} or 2) the integrated luminosity, $E_{\rm int}=\int_{\Delta t} I(t)dt$ \citep{Madiedo2014,Madiedo2019, Yanagisawa2002}.  In all cases $E_{\rm int}>E_{\rm peak}$.  According to \cite{Yanagisawa2002}, some lunar impacts have afterglows due to thermal radiation emitted from hot droplets ejected during the impact.  Those afterglows may imply that for properly estimating the energy radiated during the impact, it should be better to use $E_{\rm peak}$ instead of $E_{\rm int}$ \citep{Suggs2014}.}

\hl{The average magnitudes estimated with our data are a measure of the average intensity of the flash, namely $\langle I\rangle=\int_{\Delta t} I(t)dt/{\Delta t}$.  Therefore our luminous energy $E_r\equiv\langle I\rangle\Delta t$ (see Eq. \ref{eq:Er}) is equal to the integrated luminosity $E_{\rm int}$.  In other words, we are using here the same kind of estimation of \cite{Madiedo2014,Madiedo2019} or \cite{Yanagisawa2002}, and (for the arguments before) our luminous energy could overestimate the true radiated energy during the impact.}

\subsection{Kinetic energy}
\label{subsec:luminous}

Estimating the kinetic energy of the meteoroid from the luminous energy emitted by the impact plume is tricky.  We previously mentioned (Eq. \ref{eq:eta}) that it is customary to assume that both quantities differ only by a multiplicative ``constant'', namely the luminous efficiency $\eta$.  Although this is an oversimplification of a very complex process, \cite{Rubio2000} achieved at fitting the flux of the Leonid using a luminous efficiency of $\eta=2\EE{-3}$.  Independently \cite{Ortiz2002} and \cite{Madiedo2015a,Madiedo2015b} used the same method to obtain efficiencies in the range of $\etarange$ for different meteor showers.

If we assume that $\log\eta=\logeta$, the kinetic energy for the \LJ\ impact will be:

\beq{eq:logK}
\log \left(\frac{K}{{\rm J}}\right)=\logK
\eeq

This is equivalent to the explosion of $\Eimp$ tons of TNT.

\hl{As argued at the end of the previous section, since this kinetic energy is based on the average brightness and hence on integrated luminosity, instead of the peak luminosity, it is probably an overestimation of the actual one.  Still, the error introduced in $K$ by using $E_{\rm int}$ instead of $E_{\rm peak}$ could be compensated by the uncertainties in other factors such as $\eta$ or $f$.}

\section{Impactor and crater size}
\label{sec:sizes}

Once kinetic energy is calculated we may estimate the physical properties of the meteoroid and the crater size left by the impact.

The mass of the meteoroid can be calculated from the kinetic energy definition:

\beq{eq:mass}
M=2K/\vimp^2
\eeq

From there and assuming a proper bulk density $\rho$, we may also compute the object diameter:

\beq{eq:mass}
D=2\left(\frac{3M}{4\pi\rho}\right)^{1/3}
\eeq

The size of the crater, on the other hand, can be estimated using the well-known scaling-relationship \citep{Gault1974,Melosh1989}:

\begin{equation}
d = 0.25 \rho^{1/6} \rho_t^{−0.5} K^{0.29} (\sin\theta)^{1/3}.
\end{equation}

where d is the crater diameter and $\rho_t=\rhoRegolith$ kg/m$^{3}$ is the Moon (regolith) surface density \citep{Madiedo2015b}.

Besides $K$, the estimation of $M$, $D$ and $d$, requires educated guesses for the unknown properties $\vimp$, $\theta$ and $\rho$.

In previous works it was customary to assume  typical impact velocity between 16 - 20 km/s for sporadic meteors \citep{Brown2002,Gallant2009} or $\sim 40-72$ km/s for specific meteor showers \citep{Madiedo2014,Madiedo2015b}.  The value of $\theta$ was always guessed in the absence of observational evidence able to constrain it.

Here, our dynamical model provide us marginal probability distribution function for these quantities (see Figure \ref{fig:ppds}).  Thus, instead of replacing the value of educated guesses, we can compute {\it posterior probabilities distributions} (ppd) for the desired quantities.

For this purpose we perform a simple Monte Carlo simulation where 1,000 values of $\log K$ and the uncertain parameters $\vimp$, $\theta$ and $\rho$ were generated according to their marginal probability distribution.  The values of $\log K$ was generated assuming a Gaussian distribution of mean and standard deviation given by Eq. \ref{eq:logK}. Values of $\vimp$ and $\theta$ where independently generated using the distributions computed in Section \ref{sec:orbit}.

The case of $\rho$ is interesting.  The density of typical meteoroids impacting the Moon ranges from \rhoMin\ kg/m$^3$ (in the case of soft cometary materials) to \rhoMax\  kg/m$^3$, the density of ordinary chondrites (\citealt{Madiedo2014} and references there in).  Since we do not know the nature of the impactor, we should assume several values for the density.

For our Monte Carlo we use the following values for the density: $\rho_1=\rhoComet$ kg/m$^3$ with a probability of \pComet$\%$ (approximately matching the fraction of NEOs which are comets \footnote{\url{https://cneos.jpl.nasa.gov/stats/totals.htm}}), $\rho_2=\rhoRocky$ kg/m$^3$ with a probability of \pRocky$\%$ (arising from parent bodies with Tisserand parameters below 2, \citealt{Moorhead2017}) and $\rho_3=\rhoMetal$ kg/m$^3$ with a probability of \pMetal$\%$ (arising from parent bodies with large Tisserand parameters, \citealt{Moorhead2017}).

\begin{figure}
    \centering
    \includegraphics[width=0.45\textwidth]{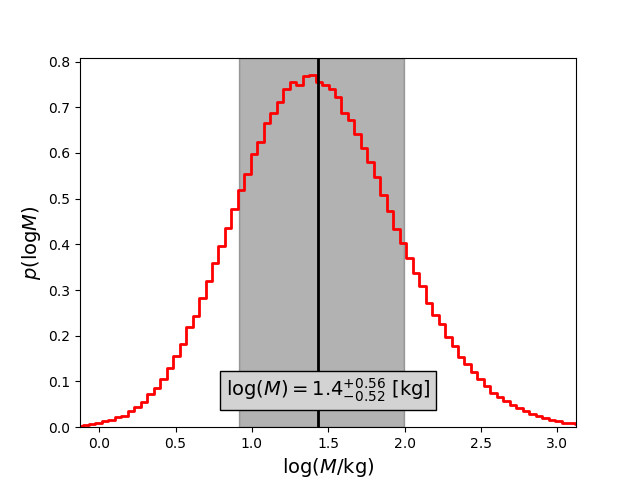}
    \includegraphics[width=0.45\textwidth]{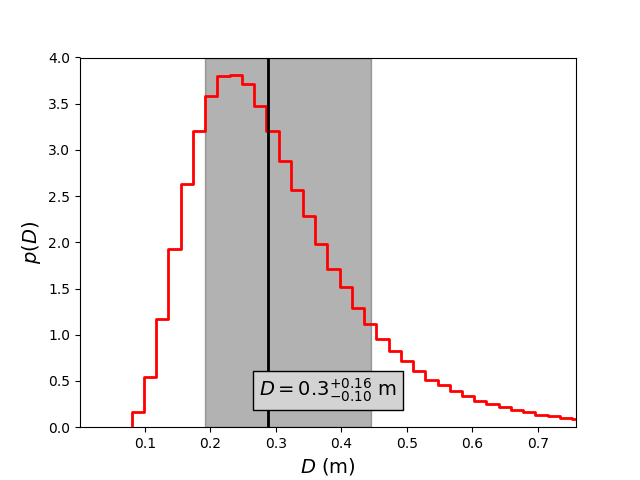}
    \includegraphics[width=0.45\textwidth]{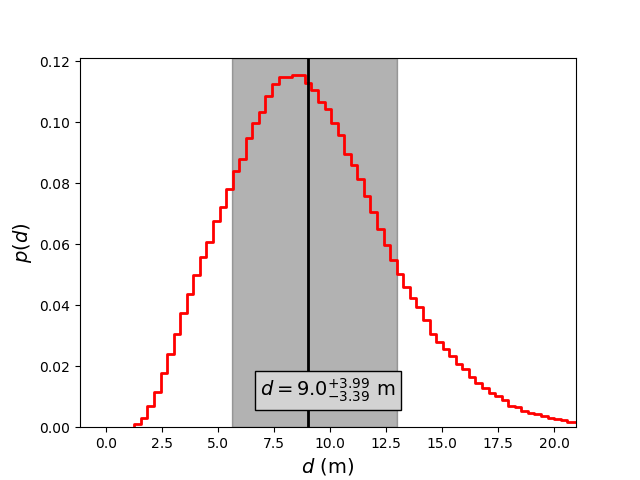}
    \caption{From top to bottom, posterior probability distributions (ppd) of meteoroid mass, diameter and crater size.}
    \label{fig:MDd}
\end{figure}

In Figure \ref{fig:MDd} we show posterior probability distributions for meteoroid mass $M$, diameter $D$ and crater size $d$.

We find that, in order to explain all the observations, the meteoroid producing the \LJ\ impact should be of the size of a ``foot ball'' \Drange\ cm diameter.  Depending on its density the mass of the object could be in the range \Mrange\ kg.  Given the shallow angle the crater diameter will be in the range between $\craterRange$ m.  This is well within the best resolution of the lunar prospecting moon satellites.

\section{Discussion}
\label{sec:disc}


The reliability of the results in this work depend on the assumptions we made in the three steps we followed: astrometry, for determining the location; astrodynamics, for estimating the speed and incoming direction of the object; and photometry, for calculating the total energy involved in the impact.

The visual alignment between the lunar image and LROC maps we used for estimating the location of the impact, is not a trivial procedure. The limited spatial resolution of the image and moon libration (not accounted in maps) were the most important sources of errors in this procedure.  Still, and as it is demonstrated by the overlapping of the visually-estimated location and the position obtained with the geometrical procedure, we are now confident that careful visual inspection could be as good as more sophisticated methods when determining the position of lunar flashes.

The almost perfect match between the position of the impact site obtained from images taken at different geographical locations and with different equipment, confirm us the reliability of the geometrical procedures devised here.  These procedures, that also demonstrate their precision at measuring the parallax of the moon, can be used in the future for collaboratively studying lunar flashes and/or performing citizen astronomy campaigns.

The most innovative aspect of this work, namely the use of GRT to constraint the kinematic properties of the impact, is also the more controversial. On one hand, since GRT depends on assuming a population from which the parent body came, their results can be considered doubtful since we ignore the orbital distribution of objects with the size of the impactor.

\hl{Assuming that the object belongs to the sporadic background implies accepting that we witnessed and incredibly rare event (there are very few sporadic meteors larger than a few centimetres).} 
\hl{Still, it is important to stress that we are not suggesting that  all objects in the size range of L1-21J in the Earth-Moon system environment are asteroids or belong to the NEOs population instead of the sporadic background.  It may also belong to the population of other specific showers.  For instance, according to a recent result published by \citealt{Clark2019}, the Taurid complex, may contain cometary fragments as large as several meters. Our aim here was to explore an alternative hypothesis to the most common scenario, namely a high speed, low density impactor.}

\hl{Conversely, if we assume that the object belongs to the NEO population, we are ignoring the fact that small NEO objects could have a different distribution than that of larger ($>$50 m) already discovered ones.}

\hl{Another limitation of GRT is the fact that it} has only been applied in restricted contexts and by only one group of authors.  The method has to be better developed and tested, both from a theoretical and an observational point of view. Our group is working in this direction but we encourage other groups to independently explore and develop the method.

\hl{In order to asses the parent population issue, we  repeated the GRT analysis of Section \ref{sec:orbit}, using as target population the sporadic meteors in the CAMS database.  For this purpose we use the orbital elements of $\sim$4000 CAMS meteoroids, having diameters larger than 2 cm.  Additionally, in order to calculate the posterior probability distributions of the key quantities, instead of assuming a mostly rocky composition, we use a density of 1000 kg/m$^3$ for 67\% of the test objects in our Monte Carlo simulation \citep{Williams2014} and 2500 kg/m$^3$ for the remaining objects.}

\hl{Our results, assuming that the object came from  the sporadic background are: $\log M= 1.3^{+0.57}_{-0.53}$ [kg], $D=0.3^{+0.19}_{-0.12}$ cm and $d=6.5^{+3.0}_{-2.5}$ m.  They are very similar  to the results of Section \ref{sec:sizes}.}

\hl{Although the distribution of speeds and incoming directions for sporadic objects are different than that of the nominal case (NEO population), the combination of all factors produce similar posterior distributions for the key quantities.  Thus, for instance, a faster sporadic meteoroid, implies a lower object mass, but since it also has a lower density, its size will be similar than that of a NEO. Consequently, we can conclude that our final results are robust against the uncertainty in the parent population of the object.}

The simple photometric procedure we devised for estimating the average G or V magnitude of lunar flashes with RGB images, provided consistent results between pictures taken with different sensors and exposure times.  This confirms that despite its simplicity, the procedure is good enough for its intended purpose.

Although, it is common that the brightness of a lunar flash be reported in terms of its peak magnitude \citep{Madiedo2015b, Xilouris2018, Avdellidou2019}, all our photometric data come from pictures having exposure times longer than flash duration.  Therefore, we cannot compare easily our average magnitude (which is a measure of the integrated brightness of the flash) with, for instance, the simple scaling laws developed by \cite{Bouley2012} for relating flash duration and peak brightness. Still, it is not hard to show that the difference between peak and average magnitude for a source having the typical light-curves of lunar flashes, it is between 2 and 3 magnitudes. Therefore, if we rely in our results, the peak brightness of the flash should be $G_{\rm peak}\sim 3.7-4.7$.

\section{Summary and conclusions}
\label{sec:conclusions}

In this paper we provide estimations of the location, orbit and energy of the meteoroid impacting the moon during the lunar eclipse of January 21, 2019.  For this purpose we use 8 RGB pictures taken by observers in an equal number of countries and 3 continents around the world.  These results arose from a timely collaboration between professional and amateur astronomers.

Although lunar impacts are constantly monitored by three major observing programs (NASA Marshall Space Flight Centre, MIDAS and NELIOTA), the \LJ\ impact happened during a total lunar eclipse and it was observed by many casual observers.  This fact provide us an unique opportunity to study this type of events using independent equipment, methods and theoretical assumptions.

Although this work uses procedures and formulas widely tested in the field, it also provides some interesting innovations.  The most important one has to be with the origin and kinematic properties of the object.  In contrast with most works in the field, we tested the hypothesis that the object came from the NEO population (mostly asteroid-related objects, in low relative velocity orbits) instead of the sporadic background (mostly cometary debris, in large relative velocity orbits).  Moreover, instead of assuming an \textit{a priori} distribution of impact speed and incoming directions, we use GRT to estimate consistently the distribution of both quantities.  Last, but not least, our results were obtained using heterogeneous images gathered by different observers around the world.  The methods developed to analyse these images could be useful in future similar collaborations or to independently test the results of professional monitoring systems.

The future discovery of the crater left by the impact (if resolvable) and its comparison with the predictions of this model, will greatly contribute to improve it and to test the validity of the
GRT technique.

As mentioned before, our work was the result of a timely collaboration between professional and amateur astronomers.  The well-known skills of amateurs to collect and process high quality data of astronomical events, together with the capacity of professionals to convert this data into scientific
results, is becoming a powerful driver of new scientific discoveries in astronomy.

\section*{Acknowledgements}

We want to thank all people in social networks that shared their
pictures and data and that finally allowed professional and amateur
astronomers to work on this.  We are particularly grateful to the
MIDAS survey for discovering the impact and announcing early without
any embargo.  This generosity allowed all of us to discover the data
hidden in pictures and videos. We especially want to thank the careful
review of the first versions of the manuscript made by Dr. Rodrigo
Leiva from Southwest Research Institute, his suggestions and comments
were very important for our work. We would also like to thank all the
amateur astronomers that contributed the images necessary for this
work: Petr Hor\'alek from Cape Verde, Gregory Hogan from USA, Fritz
Helmut Hemmerich from Canary Islands, Libor Haspl from Czech Republic,
Robert Eder Artis from Austria, Dr. Sighard Schr\"abler and Dr. Ulrike
L\"offler from Germany, and a couple more amateur astronomers who
provided their images, but were unable to meet the scientific
precision required. Last, but not least, we greatly appreciate the
insightful comments of Maria Gritsevich, Janne Sievinen and an
anonymous colleague. Their skepticism, criticism and key suggestions shaped
this work to its final form.

\bibliographystyle{mnras}

\clearpage

\section{Appendix: Lunar parallax}
\label{sec:parallax}

In addition to tangential position along the surface of the moon measured in section \ref{sec:location}, we can also measure the radial distance using lunar parallax. The idea of using lunar parallax to measure the distance to the Moon has been proposed since the time of the Ancient Greece. However, to put this into practice with high precision is often complicated by practicality. First the high dynamic range between the Moon and background stars makes it difficult to get both resolved in a single exposure. It requires coordination between multiple observers separated by vast geographical distances to be making observation simultaneously. More importantly, it requires meticulous synchronisation in the observation time among all observers.

The total lunar eclipse offer the perfect set of circumstances for lunar parallax measurements. It is an event that is observed simultaneously worldwide across vast geographical distances. The moon brightness is dimmed enough that background stars can be easily resolved. Most importantly, the meteor flash on the lunar surface guarantees that the exposures overlap to the same fractions of a second during which the impact occurred at 04:41:37 UTC on January 21, 2019.

For each image, the apparent coordinate ($\alpha',\delta'$) of the flash and centre of the Moon is measured (Table \ref{tab:locations}). The geocentric parallax correction is then calculated for each observer location. The geocentric (common) sky coordinates (\RA,\DEC) of the impact site are related to their apparent values by \citep{Duffet-Smith}:

\begin{eqnarray}
\label{eq:parallaxcorrection1}
&\tan(\alpha-\alpha') = \frac{\rho \cos \phi' \sin H}{r \cos \delta - \rho \cos \phi' \cos H}\\
\label{eq:parallaxcorrection2}
&\tan \delta' = \cos H' \left(\frac{r \sin\delta - \rho \sin\phi'}{r \cos \delta \cos H - \rho \cos \phi'}\right),
\end{eqnarray}

where $r$ is the geocentric distance in units of the (mean) Earth radius (6378.1366 km), $\rho$ is the distance of the observer from the centre of the Earth, $\phi$ is the geocentric latitude, $H$ is the geocentric hour-angle, and $H'$ is the apparent hour-angle $H' = H + \alpha'-\alpha$. The Earth profile is approximated to be an oblate spheroid with flattening ratio of 1:298.2575.

Comparing the apparent coordinates of the flash (or any other point on the surface of the moon) as measured from different observing sites, with those calculated with eq.  (\ref{eq:parallaxcorrection1}) - (\ref{eq:parallaxcorrection2}), it is possible to fit the value of $r$, $\alpha$ and $\delta$.

Using this procedure, we find that the geocentric coordinate of the impact site was $(\alpha,\delta) = (\RAgeo,\DECgeo)$ and its distance at the precise time of the event was $\Dimpgeo$ km.  A similar procedure was performed for the centre of the moon, finding $(\alpha,\delta)_{\rm centre} = (\RAcen,\DECcen)$ and a distance of $d_{\rm centre}=\Dcengeo$ km.  For comparison {\tt NASA NAIF/SPICE} predicts for the centre of the moon $(\alpha,\delta)^{\rm theo}_{\rm centre} = (\RAcenNASA,\DECcenNASA)$ and a distance $d^{\rm theo}_{\rm centre}=\DcenNASA$ km.  The difference between the theoretical predictions and those obtained with our procedure are within the errors expected for these quantities given the quality of our data.
%

With this additional information, we can provide an alternative method to calculate the selenographic coordinate of the impact using geocentric (common) sky coordinates (\RA,\DEC). With the parallax measurement of 11 lunar surface features shown in Table \ref{tab:surfacefeatures}, we are able to construct the apparent geocentric equatorial coordinates to those features and \LJ. From these apparent geocentric equatorial coordinate, we can perform the least square fit to find the transformation coefficients to transform the coordinates back to selenographic coordinates using eq. (\ref{eq:selenographictoradec1})-(\ref{eq:selenographictoradec2}) in Section \ref{subsec:geolocation}. From these transformation coefficients, the selenographic coordinate of \LJ\ can be calculated from geocentric (\RA,\DEC) to be at selenographic lat. $\latpar$ and lon. $\lonpar$.

In this parallax measurement, angle measurement remains the largest source of error. The error source from GPS in determining the observer's location is virtually negligible compared to the astrometric errors, even when assuming GPS accuracy as high as $100$ m. In all measurements we are able to obtain order of arcseconds in precision. With the median tangential distance between observers around $2000$ km and the lunar parallax angle around $0.6$ degrees between two observations, this translate to the error in distance measurements in the order of $\pm 200$ km, which is equivalent to the relative error in lunar parallax distance of 0.05\%.

\begin{figure*}
    \centering
    \includegraphics[width=0.9\textwidth]{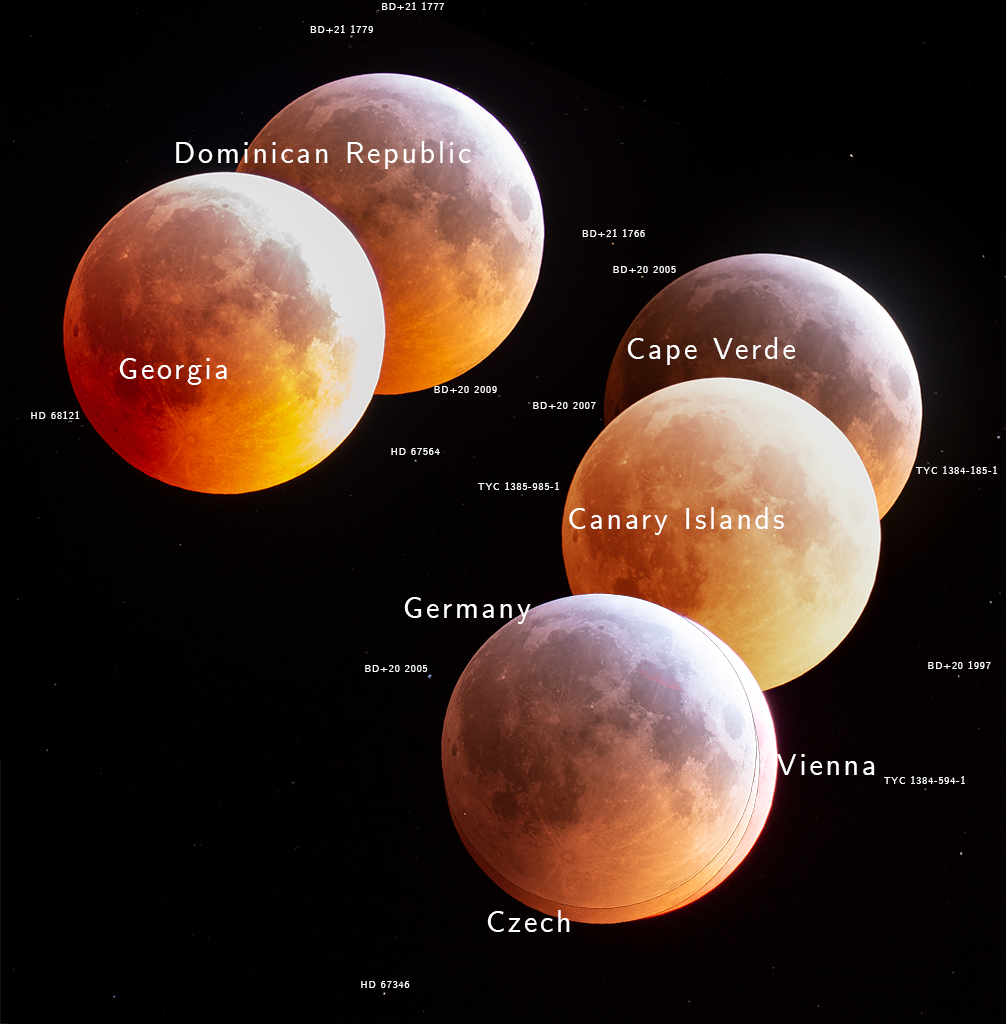}
    \caption{Overlay of all the apparent lunar position as observed by different observers. The angle differences between each images are used in Parallax calculations (section \ref{sec:parallax}). All the images are rearranged so that the background stars are alligned. Some reference stars used in photometry are displayed. The apparent positions trace a rough geographical location of the observers, with observations from Europe stacked on the bottom right, the Macaronesian (East Atlantic) islands on the right, with Continenal US and the Caribbean on the top left.}
    \label{fig:parallax}
\end{figure*}

The precision provided by this method is enough for the distances between different selenographic coordinates to be visible (see Table \ref{tab:surfacefeatures}). The diurnal libration is noticeable among images that results in differing amount of parallax between the centre and near the limb of the moon surface. This resulted in parallax distances that are different among different position along the lunar surface, with selenographic coordinate closer to the centre having shorter distance. These variation in distances among different selenographic coordinates is consistent with subtracting Earth-Moon distance with spherical projection on the surface of the moon with radius $R_{moon} = 1,737.5$ km. This is comparable to having a depth perception that allow for the curvature of the moon to be perceived via parallax. In fact, by using a cross-eyed technique on a pair of lunar images in Figure \ref{fig:parallax} one can easily see the eclipsing moon in 3D.

\begin{table*}
\centering
\resizebox{\textwidth}{!}{%
\begin{tabular}{c|cc|rrr|rrr|rr}
\hline\hline
Star name & 
\multicolumn{2}{c|}{Sky Position$^\dagger$} & 
\multicolumn{3}{c|}{Magnitudes} & 
\multicolumn{3}{c|}{Counts (AP$^{\dagger\dagger}$)} & 
\multicolumn{2}{c}{Coordinates}\\
 & RA & DEC & B & V & G$^\ddagger$ & Red & Green & Blue & X & Y\\
\hline

\hline
\multicolumn{11}{c}{RD}\\
\hline
BD+21 1766 & 8.14149 & 20.59967 & 10.69 & 9.03 & 8.35 & 14161 & 6726 & 10608 & 4138 & 1119\\
BD+20 2007 & 8.14449 & 20.28416 & 10.52 & 9.82 & 9.79 & 6129 & 8166 & 5846 & 4020 & 2589\\
HD 67564 & 8.15532 & 20.11876 & 9.38 & 9.09 & 9.02 & 8175 & 13559 & 8294 & 3356 & 3389\\
TYC 1385-899-1 & 8.19466 & 20.30565 & 11.34 & 10.22 & 10.01 & 4051 & 4038 & 4235 & 752 & 2652\\
BD+21 1779 & 8.18160 & 20.77903 & 10.20 & 9.22 & 8.95 & 9724 & 9539 & 9033 & 1494 & 423\\
BD+21 1777 & 8.17964 & 20.83795 & 11.11 & 10.18 & 9.68 & 7304 & 4911 & 5530 & 1609 & 145\\
BD+20 2009 & 8.14832 & 20.27791 & 10.82 & 10.52 & 10.44 & 3494 & 6623 & 4223 & 3772 & 2630\\
BD+20 2005 & 8.13657 & 20.56272 & 10.35 & 9.04 & 8.63 & 11959 & 7866 & 9402 & 4467 & 1274\\
TYC 1385-939-1 & 8.19105 & 20.27807 & 12.78 & 10.66 & 10.35 & 4282 & 1768 & 2955 & 994 & 2768\\
L1-21J & 8.1826* & 20.2841* & - & - & - & 1621 & 1519 & 1619 & 1562 & 2712\\

\hline
\multicolumn{11}{c}{Georgia}\\
\hline
TYC 1385-116-1 & 8.17401 & 20.60905 & 11.56 & 10.81 & 10.58 & 42396 & 48328 & 43515 & 1416 & 1255\\
HD 68121 & 8.19736 & 19.96890 & 10.03 & 9.64 & 9.56 & 11381 & 158217 & 167910 & 3346 & 2285\\
TYC 1385-1052-1 & 8.17914 & 19.83440 & 11.55 & 10.26 & 10.01 & 71357 & 78914 & 60751 & 3769 & 1512\\
TYC 1385-368-1 & 8.16140 & 20.02100 & 12.30 & 11.71 & 10.94 & 30182 & 34984 & 31206 & 3213 & 742\\
TYC 1385-188-1 & 8.19596 & 20.50482 & 10.91 & 10.87 & 10.51 & 45118 & 68999 & 84704 & 1717 & 2198\\
TYC 1385-1391-1 & 8.19691 & 19.78950 & 12.80 & 11.54 & 11.06 & 26349 & 28871 & 24964 & 3893 & 2276\\
TYC 1385-1610-1 & 8.19571 & 19.64252 & 11.66 & 11.22 & 11.22 & 21978 & 30030 & 33100 & 4340 & 2233\\
TYC 1385-1675-1 & 8.20378 & 19.69782 & 11.78 & 10.60 & 9.89 & 66890 & 69991 & 55405 & 4166 & 2577\\
TYC 1385-376-1 & 8.20688 & 19.68792 & 12.17 & 11.28 & 11.39 & 18135 & 24503 & 26564 & 4194 & 2710\\
L1-21J & 8.1968* & 20.0198* & - & - & - & 42672 & 29831 & 19853 & 3192 & 2255\\

\hline
\multicolumn{11}{c}{CapeVerde}\\
\hline
BD+20 2005 & 8.13657 & 20.56272 & 10.35 & 9.04 & 8.63 & 17763 & 18461 & 9372 & 2633 & 469\\
BD+21 1766 & 8.14149 & 20.59967 & 10.69 & 9.03 & 8.35 & 16788 & 17315 & 9183 & 2697 & 251\\
BD+20 2007 & 8.14449 & 20.28416 & 10.52 & 9.82 & 9.79 & 7700 & 7629 & 4055 & 1776 & 320\\
TYC 13841851 & 8.09468 & 20.42690 & 10.93 & 10.06 & 9.90 & 7830 & 7740 & 4465 & 2604 & 2219\\
TYC 13843851 & 8.12374 & 20.08094 & 12.55 & 11.70 & 11.24 & 1944 & 1520 & 1097 & 1375 & 1271\\
TYC 13859851 & 8.14128 & 20.12704 & 10.57 & 10.13 & 10.04 & 6518 & 6032 & 3268 & 1356 & 542\\
BD+20 2009 & 8.14832 & 20.27791 & 10.82 & 10.52 & 10.44 & 4402 & 4829 & 2330 & 1726 & 171\\
TYC 13845091 & 8.11375 & 20.03512 & 11.79 & 11.63 & 11.50 & 1506 & 1586 & 814 & 1330 & 1697\\
HD 67150 & 8.12391 & 19.81766 & 8.27 & 7.69 & 7.54 & 56496 & 56205 & 30291 & 624 & 1424\\
L1-21J & 8.1297* & 20.2223* & - & - & - & 2708 & 2535 & 1816 & 1726 & 950\\

\hline
\multicolumn{11}{c}{CanaryIslands}\\
\hline
HD 67150 & 8.12391 & 19.81766 & 8.27 & 7.69 & 7.54 & 271150 & 348264 & 236429 & 3884 & 2146\\
HD 67424 & 8.14424 & 19.77072 & 8.53 & 8.48 & 8.47 & 93949 & 154985 & 139900 & 3825 & 2801\\
TYC 1385-985-1 & 8.14128 & 20.12704 & 10.57 & 10.13 & 10.04 & 30721 & 41212 & 29098 & 3069 & 2508\\
HD 67564 & 8.15532 & 20.11876 & 9.38 & 9.09 & 9.02 & 71428 & 99635 & 76508 & 2975 & 2946\\
BD+20 2007 & 8.14449 & 20.28416 & 10.52 & 9.82 & 9.79 & 37012 & 47327 & 31202 & 2700 & 2519\\
BD+20 2009 & 8.14832 & 20.27791 & 10.82 & 10.52 & 10.44 & 19677 & 28174 & 21551 & 2683 & 2640\\
BD+20 2005 & 8.13657 & 20.56272 & 10.35 & 9.04 & 8.63 & 108293 & 105234 & 47290 & 2152 & 2118\\
TYC 1384-185-1 & 8.09468 & 20.42690 & 10.93 & 10.06 & 9.90 & 33690 & 37133 & 21446 & 2780 & 906\\
TYC 1384-1748-1 & 8.10977 & 19.80182 & 11.72 & 11.26 & 11.13 & 9836 & 12076 & 8110 & 4030 & 1719\\
L1-21J & 8.1291* & 19.9900* & - & - & - & 138565 & 94635 & 30931 & 3465 & 2210\\

\hline
\multicolumn{11}{c}{\it Continues in Table \ref{tab:referencestars2}}\\
\hline\hline
\multicolumn{11}{l}{\footnotesize $^\ddagger$ J2000, $^\ddagger$ Gaia G-magnitude, $^{\dagger\dagger}$ Aperture photometry}\\
\multicolumn{11}{l}{\footnotesize $^*$ Calculated coordinates (see Section \ref{sec:parallax})}\\
\end{tabular}}
\caption{Reference stars properties, photometry and astrometry results for the pictures analysed in this work.\label{tab:referencestars1}}
\end{table*}
\begin{table*}
\centering
\resizebox{\textwidth}{!}{%
\begin{tabular}{c|cc|rrr|rrr|rr}
\hline\hline
Star name & 
\multicolumn{2}{c|}{Sky Position$^\dagger$} & 
\multicolumn{3}{c|}{Magnitudes} & 
\multicolumn{3}{c|}{Counts (AP$^{\dagger\dagger}$)} & 
\multicolumn{2}{c}{Coordinates}\\
 & RA & DEC & B & V & G$^\ddagger$ & Red & Green & Blue & X & Y\\
\hline
\hline
\multicolumn{11}{c}{\it Continued from Table \ref{tab:referencestars1}}\\

\hline
\multicolumn{11}{c}{Germany}\\
\hline
HD 67424 & 8.14424 & 19.77072 & 8.53 & 8.48 & 8.47 & 2978 & 6118 & 5076 & 1358 & 950\\
HD 67346 & 8.13574 & 19.21594 & 8.23 & 7.63 & 7.48 & 5117 & 10885 & 6747 & 918 & 2278\\
TYC 1384-1330-1 & 8.09963 & 19.38749 & 11.00 & 10.05 & 9.76 & 704 & 1086 & 785 & 2221 & 2542\\
TYC 1384-594-1 & 8.08099 & 19.88256 & 10.66 & 9.87 & 9.61 & 744 & 1538 & 913 & 3395 & 1815\\
BD+20 1997 & 8.08191 & 20.09054 & 10.21 & 9.36 & 9.09 & 986 & 2104 & 1243 & 3628 & 1357\\
BD+20 2002 & 8.11703 & 20.30866 & 10.83 & 10.37 & 10.20 & 464 & 969 & 671 & 2841 & 282\\
TYC 1385-985-1 & 8.14128 & 20.12704 & 10.57 & 10.13 & 10.04 & 557 & 1244 & 854 & 1890 & 247\\
TYC 1384-963-1 & 8.12454 & 19.45866 & 12.73 & 11.52 & 10.75 & 120 & 282 & 244 & 1562 & 1955\\
TYC 1385-1668-1 & 8.13691 & 19.31428 & 11.70 & 11.37 & 11.08 & 213 & 421 & 232 & 1008 & 2045\\
L1-21J & 8.1342* & 19.5625* & - & - & - & 3699 & 5351 & 2730 & 1400 & 1568\\

\hline
\multicolumn{11}{c}{Czech}\\
\hline
HD 67346 & 8.13574 & 19.21594 & 8.23 & 7.63 & 7.48 & 5081 & 6300 & 6616 & 1341 & 2032\\
HD 67424 & 8.14424 & 19.77072 & 8.53 & 8.48 & 8.47 & 3480 & 4637 & 6015 & 2871 & 992\\
TYC 1385-985-1 & 8.14128 & 20.12704 & 10.57 & 10.13 & 10.04 & 1215 & 1325 & 1218 & 3997 & 685\\
BD+20 2007 & 8.14449 & 20.28416 & 10.52 & 9.82 & 9.79 & 1366 & 1473 & 1508 & 4416 & 358\\
BD+20 2002 & 8.11703 & 20.30866 & 10.83 & 10.37 & 10.20 & 951 & 1034 & 1041 & 4964 & 1493\\
HD 66551 & 8.08129 & 20.23307 & 9.07 & 8.93 & 8.91 & 2060 & 2469 & 3168 & 5358 & 3101\\
BD+20 1997 & 8.08191 & 20.09054 & 10.21 & 9.36 & 9.09 & 1650 & 1688 & 1666 & 4917 & 3250\\
TYC 1384-594-1 & 8.08099 & 19.88256 & 10.66 & 9.87 & 9.61 & 1113 & 1399 & 1142 & 4308 & 3544\\
TYC 1384-1330-1 & 8.09963 & 19.38749 & 11.00 & 10.05 & 9.76 & 1102 & 869 & 846 & 2488 & 3362\\
L1-21J & 8.1334* & 19.5386* & - & - & - & 6324 & 5347 & 3548 & 2360 & 1746\\

\hline
\multicolumn{11}{c}{Vienna}\\
\hline
HD 67346 & 8.13574 & 19.21594 & 8.23 & 7.63 & 7.48 & 31494 & 42730 & 25476 & 1570 & 2847\\
HD 67424 & 8.14424 & 19.77072 & 8.53 & 8.48 & 8.47 & 12565 & 19687 & 16399 & 1628 & 1493\\
TYC 13859851 & 8.14128 & 20.12704 & 10.57 & 10.13 & 10.04 & 2228 & 3462 & 2078 & 1939 & 695\\
BD+20 2007 & 8.14449 & 20.28416 & 10.52 & 9.82 & 9.79 & 3222 & 4585 & 2610 & 1929 & 304\\
TYC 13841851 & 8.09468 & 20.42690 & 10.93 & 10.06 & 9.90 & 3633 & 4292 & 2017 & 3635 & 389\\
BD+20 1997 & 8.08191 & 20.09054 & 10.21 & 9.36 & 9.09 & 7832 & 8560 & 4440 & 3851 & 1274\\
TYC 13845941 & 8.08099 & 19.88256 & 10.66 & 9.87 & 9.61 & 4928 & 5578 & 3010 & 3758 & 1762\\
TYC 138413301 & 8.09963 & 19.38749 & 11.00 & 10.05 & 9.76 & 3260 & 4414 & 2270 & 2855 & 2752\\
HD 67564 & 8.15532 & 20.11876 & 9.38 & 9.09 & 9.02 & 6683 & 10083 & 7036 & 1476 & 594\\
L1-21J & 8.1315* & 19.5562* & - & - & - & 8457 & 4877 & 2342 & 1921 & 2097\\

\hline\hline
\end{tabular}}
\caption{Continuation of Table \ref{tab:referencestars1}.\label{tab:referencestars2}}
\end{table*}

\bsp 
\label{lastpage}
\end{document}